\documentclass[%
 aip,
rsi,%
 amsmath,amssymb,
reprint,%
floatfix]{revtex4-1}

\usepackage{graphicx}
\usepackage{dcolumn}
\usepackage{bm}
\usepackage{subfigure}

\newcommand*{\cD}{\mathcal{D}}

\newcommand*{\cI}{\mathcal{I}}

\newcommand*{\ep}{\epsilon}

\newcommand*{\B}{\bm{B}}

\newcommand*{\J}{\bm{J}}

\newcommand*{\dl}{\bm{\nabla}}

\newcommand*{\BD}{\bm{B}\cdot\bm{\nabla}}

\newcommand*{\lbr}{\left(}
\newcommand*{\rbr}{\right)}

\newcommand{\iotabar}{\mbox{$\iota\!\!$-}}

\newcommand*{\alphabar}{\overline{\alpha}}

\makeatletter
\newsavebox{\@brx}
\newcommand{\llangle}[1][]{\savebox{\@brx}{\(\m@th{#1\langle}\)}%
  \mathopen{\copy\@brx\mkern2mu\kern-0.9\wd\@brx\usebox{\@brx}}}
\newcommand{\rrangle}[1][]{\savebox{\@brx}{\(\m@th{#1\rangle}\)}%
  \mathclose{\copy\@brx\mkern2mu\kern-0.9\wd\@brx\usebox{\@brx}}}
\makeatother

\begin{document}

\preprint{AIP/123-QED}

\title[Exact closed line 3D vacuum magnetic  field]{Exact non-symmetric closed line vacuum magnetic fields in a topological torus}

\author{Harold Weitzner}
\email{weitzner@cims.nyu.edu}
\author{Wrick Sengupta}%
\affiliation{ 
Courant Institute of Mathematical Sciences, New York University, New York, New York 10012, USA
}%


\begin{abstract}
Non-symmetric vacuum magnetic fields with closed magnetic field lines are of interest in the construction of stellarator equilibria. Beyond the result of D.Lortz (ZAMP \textbf{21}, 196 (1970)), few results are available. This work presents a closed-form expression for a class of vacuum magnetic fields in a topological torus with closed field lines. We explicitly obtain the invariants of such a field. We finally show that a three-dimensional low beta magnetohydrodynamic (MHD) equilibrium may be constructed in a topological torus, starting with these closed line vacuum magnetic fields.
\end{abstract}

\keywords{closed line, non-symmetric, three-dimensional, vacuum magnetic fields, exact solution}
\maketitle

\section{Introduction}
The existence of magnetic fields for non-dissipative plasma models in a multiply connected domain with nested flux surfaces or exclusively closed field lines remains an incompletely understood topic. While the existence of magnetic surfaces is easily demonstrated in domains with continuous symmetry such as toroidal, cylindrical, helical, or translational, it is well understood that such surfaces are easily destroyed by small perturbations that destroy the original symmetry. In particular, if we look for continuously deformable smooth and continuous non-symmetric solutions of ideal MHD, the pressure and rotation transform profiles cannot be arbitrarily chosen to avoid the singular divisors on rational surfaces \citep{grad1967toroidal,newcomb1959magnetic,hudsonKrauss20173D_cont_B}. 

To avoid magnetic resonances, modern stellarators like W7-X and HSX are designed to have low global magnetic shear. The behavior of low-shear magnetic field systems can be markedly different in the neighborhood of closed field lines from generic ergodic surfaces \citep{firpo2011study, faber2018stellarator}. Numerous experimental results from Wendelstein VII-A/AS \citep{hirsch2008majorW7AS,brakel2002energytransp_rational_iota_W7AS,brakel1997W7AS_EB_shear} and numerical results \citep{wobig1987localized_pert_w7A,andreevaW7Xvacuum} support the idea that optimum confinement is usually found close to certain low-order rational surfaces. For small but arbitrary perturbations, the islands on these surfaces are
found \citep{wobig1987localized_pert_w7A} to be exponentially small in size consistent with Arnold's theorem on ``properly degenerate systems." The stability advantages of low-shear MHD systems compared to the high shear systems have also been discussed \citep{grad1973magnetofluid,strauss1981limiting_beta_vacuum}. W7-X has two very different configurations, which both access regions close to $\iotabar=1$, namely, the standard configuration and the high-$\iotabar$ configuration \citep{andreevaW7Xvacuum,klinger2019overview}. In the former case, shear is relatively large compared to the latter, and there are islands near $\iotabar=1$ in the former but not in the latter. Therefore, it is indeed possible to have low-shear regions near low-order rational surfaces under experimental conditions, but careful design is needed \citep{lazerson2019tuning}.

The possibility of the existence of nested surfaces or closed field lines in generic three-dimensional domains is an open mathematical question with at least one notable example of an ideal magnetohydrodynamic equilibrium in a non-symmetric domain as given by D.Lortz \cite{lortz1970existenz}. Lortz showed that in a system with reflection symmetry, a non-symmetric three-dimensional MHD equilibrium with a smooth pressure profile can be constructed iteratively starting with a vacuum field which has closed field lines and zero shear. A significant drawback of this result is that the rotation transform for such an equilibrium is zero. 

The difficulties in obtaining exact solutions or even perturbative analytic expressions for three-dimensional non-symmetric vacuum magnetic fields with surfaces have been pointed out \citep{cary1982vacuum,freidberg_idealMHD, sengupta_weitzner2018}. For ideal magnetohydrodynamics, the literature has a rich collection of formal expressions of equilibria, expansions being in the amplitude of magnetic field components \cite{weitzner2014ideal,cary1984construction} or in the distance from a magnetic axis \cite{Mercier1964, Solovev1970}, a magnetic line or a magnetic surface \cite{weitzner2016expansions,sengupta_weitzner_2019}.  
Some of these expansions are shown to be able to be carried to all orders; many are carried out only to the first few orders. The convergence of these series expansions is not known, nor is their non-convergence. In an earlier paper \cite{sengupta_weitzner_2019}, we treated the formally simple problem of the expansion of a vacuum magnetic field, the expansion being in the distance of a planar surface, say $x=0$. The magnetic field was found to be periodic of period $2\pi$ in the two orthogonal coordinates $y$ and $z$. Such periodicity renders the domain of definition of the magnetic field doubly connected, i.e., a ``torus. " We developed expansions to all orders in the distance $x$ from the plane $x=0$. The ability to expand to all orders, at least as we carried it out, severely limited the possible magnetic fields. We do not know if these limitations are real or a consequence of our approach to the problem. We did not examine the convergence of the series, either. 

In this work, we return to the problem previously studied and extend the analysis further. In section \ref{sec:construnction}, we take the magnetic field structure used there, we extend the scalar potential $\Phi$ for the vacuum field to an explicit representation in all space, and finally show that in some open set containing the plane $x=0$ every magnetic field line is closed. We comment on a few properties of such fields. The representation of $\Phi$ allows either finite or infinite series representations. We omit the restrictions necessary in the latter case. In section \ref{sec:model_vacuum}, we treat one particularly simple representation of a closed line system in greater detail and obtain the Clebsch potentials. In section \ref{sec:Lortz}, we show that the Lortz analysis allows this special vacuum field to be the starting point for the expansion in power series of plasma beta $\beta\equiv p/B^2$. We discuss how the Lortz iteration scheme allows us to construct non-symmetric MHD equilibrium with closed field lines. We require only a minimal result from the earlier work. 

\section{Construction of a vacuum magnetic field with closed field lines \label{sec:construnction}}
Since $\B$ is a vacuum magnetic field, we introduce a scalar potential $\Phi(x,y,z)$, such that
\begin{align}
    \B =\dl \Phi.
\end{align}
We assume that $\Phi$ is an even function of $x$ and that on the plane $x=0$,
\begin{align}
    \Phi(0,y,z)=\mu(y)+\mu(z).
    \label{phi0form}
\end{align}
Here $\mu(w)$ is an odd function of $w$ of the form 
\begin{align}
    \mu(w)= w +\ep \tilde{\mu}(w),
\end{align}
where $\tilde{\mu}(w)$ is periodic of period $2\pi$ in $w$ and $\ep$ is a constant. We found \cite{sengupta_weitzner_2019} for such fields that one could find magnetic surfaces to all orders. The properties that $\Phi(x,y,z)$ be even in $x$, a solution of Laplace's equation and the form (\ref{phi0form}) completely characterizes $\Phi$ and we obtain
\begin{align}
    \Phi(x,y,z)= \frac{1}{2} &\left[ \mu(y+ix)+\mu(y-ix) \right.\nonumber\\
    & +\left. \mu(z+ix)+\mu(z+ix) \right].
\end{align}
If we then use the structure and parity of $\mu$ we conclude that
\begin{align}
    \Phi(x,y,z)=& (y+z) \nonumber\\
    +\frac{1}{2}\ep \sum_{m=1}^{\infty} & \frac{a_m}{m}\left( \sin{m (y+i x)}+\sin{m (y+i x)}\right. \nonumber\\
    &+\left. \sin{m (z+i x)}+\sin{m (z+i x)}\right), \nonumber 
\end{align}
or,
\begin{align}
    \Phi= (y+z)+\ep \sum_{m=1}^{\infty} & \frac{a_m}{m}\lbr \sin{m y}+\sin{m z}\rbr\cosh{m x}.
    \label{Phixyzform}
\end{align}
We conclude immediately from the form (\ref{Phixyzform}) that the coefficients $a_m$ must be severely constrained so that the series represents an analytic function of $x$ in some domain $|x|<x_{\text{max}}$. Such conditions are lost in the formal series analysis used previously. For simplicity we assume here that at most a finite number of coefficients $a_m$ are non-zero and that the parameter $\ep$ is small enough such that
\begin{align}
    1 \geq L >\sum_{m=1}^{\infty} |a_m|
    \label{ep_ineq}
\end{align}
We could extend the analysis to allow infinitely non-zero values of $a_m$, provided more stringent inequalities of the form (\ref{ep_ineq}) were imposed. With $\Phi$ as given by (\ref{Phixyzform}) satisfying (\ref{ep_ineq}), we now show that every magnetic field line in some region is closed. 

We observe that a magnetic field line satisfies the equation 
\begin{align}
\dfrac{d\bm{x}}{dt}= \B = \dl \Phi
    \label{BODE}
\end{align}
It is convenient to change the independent coordinates from $(x,y,z)$ to $(X,Y,Z)$ where
\begin{subequations}
\begin{align}
x&= \sqrt{2}X, \\  y&=\sqrt{2}\frac{1}{\sqrt{2}}\lbr Y+Z\rbr \\z&=\sqrt{2}\frac{1}{\sqrt{2}}\lbr Z-Y\rbr .
\end{align}
\label{XYZ}    
\end{subequations}
The transformation (\ref{XYZ}) is a rotation of $(\pi/4)$ about the $x$ axis followed by a stretching of all coordinates by a factor $\sqrt{2}$. Clearly, we may follow magnetic field lines in either coordinates to demonstrate closure, as the relation (\ref{BODE}) is only modified by a coordinate stretching. We find easily that 
\begin{align}
    \Phi/2 =  Y + \ep \sum_{m=1}^{\infty} \frac{a_m}{m}\sin{m Y}\cos{m Z} \cosh{\lbr \sqrt{2}m X \rbr}.
    \label{Phi_series}
\end{align}
Thus, the equations for a magnetic field line are
\begin{subequations}
\begin{align}
    \dfrac{dX}{d\ell}=& \sqrt{2} \ep \sum_{m=1}^{\infty} a_m \sin{m Y}\cos{m Z} \sinh{\lbr \sqrt{2}m X \rbr} \label{Xdot}\\
    \dfrac{dY}{d\ell}=& 1+ \ep \sum_{m=1}^{\infty} a_m \cos{m Y}\cos{m Z} \cosh{\lbr \sqrt{2}m X \rbr} \label{Ydot}\\
    \dfrac{dZ}{d\ell}=& -\ep \sum_{m=1}^{\infty} a_m \sin{m Y}\sin{m Z} \cosh{\lbr \sqrt{2}m X \rbr}. \label{Zdot}
\end{align}
\end{subequations}
We note that there is a unique magnetic field through any given point $(X_0,Y_0,Z_0)$. Further, the field line is an analytic function of the initial values, so that in any bounded domain the field is a uniformly continuous function of the initial data. Next we observe that any field line through $X_0=0$ remains in the $X=0$ plane. Equally, every field line through the $Z= M \pi$ plane, remains in that plane for every integer $M$. Finally, we conclude that there is an open set containing the section of the plane $X_0=0$ with $|X|<X_{\text{max}}, |Y|<3 \pi,|Z|<3 \pi$ such that
\begin{align}
    1 > \ep \sum_{m=1}^{\infty} |a_m|\sinh{\sqrt{2}m X_{\text{max}}}
    \label{ep_bound},
\end{align}
and every field line remains in that domain. In view of (\ref{Ydot}), $Y(\ell)$ is a monotonic increasing function of $\ell$. We may therefore write the equations in the form
\begin{subequations}
\begin{align}
    \dfrac{dX}{dY}&= \sqrt{2}\ep \frac{\sin Y}{\cD} \sum_{m=1}^{\infty} a_m \frac{\sin{m Y}}{\sin Y}\cos{m Z}\sinh{(\sqrt{2}m X)}\\
    \dfrac{dZ}{dY}&= -\ep \frac{\sin Y}{\cD} \sum_{m=1}^{\infty} a_m \frac{\sin{m Y}}{\sin Y}\sin{m Z}\sinh{(\sqrt{2}m X)}\\
    \cD &= 1 + \ep\sum_{m=1}^{\infty} a_m \cos{m Y}\cos{m Z}\cosh{(\sqrt{2}m X)}.
\end{align}
\label{BxBzODE}
\end{subequations}
If we change the independent variables from $Y$ to $\mu= \cos{Y}$, then as $Y$ ranges from 0 to $2\pi$, $\mu$ ranges from $1$ to $-1$ and back to $1$. Since, $X(0)=X(2\pi)$ and $Z(0)=Z(2\pi)$, the field line is closed. 

The closure of the field lines found here for the field with scalar potential $\Phi$ depends on the reflection symmetry that $\Phi$ be even in $X$ and odd in $Y$, essentially the same discrete symmetry used by Lortz. We append a standard but rarely discussed result for our system: that the rotational transform of a closed line magnetic field is not well defined. We start from the field as given by (\ref{ep_bound},\ref{BxBzODE}). We may generalize the transformation \ref{XYZ} to
\begin{subequations}
\begin{align}
X'&= \sqrt{M^2+N^2}X, \\  Y'&=\lbr M Y+N Z\rbr \\Z'&=\lbr M Z-N Y\rbr,
\end{align}
\label{XYZp}    
\end{subequations}
\noindent{\hspace{-0.08in}}where $M,N$ are relatively prime. Again, (\ref{XYZp}) is a rotation about the $X$ axis by an angle of $\arctan{(M/N)}$ followed by a stretch. Thus, again, the field lines are periodic of period $2\pi$ in $Y$. The field line starting at $Y=0, Z=z_0$ after one period $Y=2\pi$, has $Z=z_0$. Correspondingly, the two points are $(Y',Z')=(0,N z_0)$ and $(2\pi M, -2\pi N+M z_0)$. Hence, the field lines will have a twist of $M/N$ or $N/M$. Usually, one identifies the twist with the rotational transform, but clearly, the identification here is spurious. Thus, the concept of a rotation transform, in this case, is ill-defined.

Finally, we note that while periodicity in the expression for $\Phi$ in the variable $Z$ is essential, the specific form for $\Phi$  as a function of $Z$ is not.  Thus, each term $\cos(m Z)$ might be replaced by $\cos(m Z+\delta_m)$, and the subsequent analysis would be possible with only minor modifications at each step. These modifications allow a more extensive range of vacuum fields and of equilibria than are shown explicitly.

\section{Clebsch representation for the vacuum magnetic field}
\label{sec:model_vacuum}
It is useful to explore a simple, explicit version of these results. To that end we set $a_1=\ep, a_m=0$ for all $m>1$ i.e.
\begin{align}
    \varphi = Y+\ep\sin{Y}\cos{Z}\cosh(\sqrt{2}X).
    \label{modelPhi}
\end{align}
where $\varphi=\Phi/2$. We shall now construct $\psi,\alpha$ explicitly such that 
\begin{align}
    \dl \varphi = \dl \psi \times \dl \alpha,
    \label{phi_psi_alpha}
\end{align}
using method of characteristics. We solve
\begin{align}
    \frac{dX}{B_X}=\dfrac{dY}{B_Y}=\dfrac{dZ}{B_Z} \label{MoC}  \\
    \frac{dX}{\ep\sqrt{2}\sinh{(\sqrt{2}X)}\cos{Z}\sin{Y}}=& \dfrac{dZ}{-\ep\cosh{(\sqrt{2}X)}\sin{Z}\sin{Y}} \nonumber 
     \\
    =\dfrac{dY}{1+\ep\cosh{(\sqrt{2}X)}\cos{Z}\cos{Y}}. \nonumber
\end{align}
We observe that the equations are invariant under 
\begin{align}
    X\rightarrow -X, \:\: Y\rightarrow -Y ,\:\: Z\rightarrow -Z.
    \label{parityXYZ}
\end{align}
Hence, the invariants must be even functions of all the coordinates. 
Since $B_x$ and $B_z$ vanish at $X=0$ and $Z=0$ respectively, the characteristics have to be calculated separately on the $X=0$ and $Z=0$ planes and for $X\neq 0, Z\neq 0$ regions. We now discuss each of these regions separately.

\subsubsection{Invariants on the $X=0$ plane}
Substituting $X=0$ in (\ref{MoC}) we obtain
\begin{align}
   \dfrac{dZ}{-\ep\sin{Z}\sin{Y}} =\dfrac{dY}{1+\ep\cos{Z}\cos{Y}}. \label{MoC_X0}
\end{align}
Since $B_z$ vanishes at $Z=0$, field lines do not cross the $Z=0$ plane. Therefore, we treat $Z>0$ and $Z<0$ regions separately. For $Z\neq0$, we can easily integrate (\ref{MoC_X0}) subject to parity condition (\ref{parityXYZ}) and obtain
\begin{align}
    \cI_x \sin{|Z|}-\cos{Z}=\ep \cos{Y},
    \label{Ix}
\end{align}
where, $\cI_x$ is a constant along the field lines. Thus, $X=0$ and $\cI_x=$ constant, are the two invariants in this case. We note that for small $\ep$, $\cI_x\approx \cot{|Z|}$ and that (\ref{Ix}) shows $\cI_x$ is singular at $Z=0,\pm \pi$.

\subsubsection{Invariants on the $Z=0$ plane}
For $Z=0$, (\ref{MoC}) simplifies to
\begin{align}
    \frac{dX}{\ep\sqrt{2}\sinh{(\sqrt{2}X)}\sin{Y}}=\dfrac{dY}{1+\ep\cosh{(\sqrt{2}X)}\cos{Y}}. \label{MoC_Z0}
\end{align}
Using the parity condition  (\ref{parityXYZ}), we integrate (\ref{MoC_Z0}) and get
\begin{align}
 \cI_Z&=\ep \sqrt{\sinh{(\sqrt{2}|X|})}\cos{Y} +\int^{|X|}_{0}\dfrac{dX'}{\sqrt{2\sinh{(\sqrt{2}X')}}},
    \label{Iz}
\end{align}
with $\cI_Z$ being the invariant along with $Z=0.$ For small values of $\ep$, $\cI_Z$ is given by the second term.

\subsubsection{Invariants when $X\neq 0, Z\neq 0$}
Finally, when $X\neq 0, Z\neq 0$, it is easily verified that the following two functions are constants along any magnetic field satisfying (\ref{MoC}) together with (\ref{parityXYZ}):
\begin{subequations}
\begin{align}
    \psi(X,Y,Z)&=\sqrt{2\sinh{(\sqrt{2}|X|})}\sin{|Z|}\label{psi_exp}\\
    \alpha(X,Y,Z)&=\ep \sqrt{\sinh{(\sqrt{2}|X|})}\cos{Y}\label{alpha_exp}\\ &+\sigma_z\int^{|X|}_{|X_0|}\dfrac{dX'}{\sqrt{2\sinh{(\sqrt{2}X')}-\psi^2}} ,\nonumber
\end{align}
\label{psi_alpha}
\end{subequations}

where,
 \begin{align}
     \sigma_z= \text{sign}(\cos{Z}),\: X_0=
 \dfrac{1}{\sqrt{2}}\sinh^{-1}{\lbr\frac{\psi^2}{2}\rbr}.
 \end{align}
In obtaining the above we have made use of the identity
\begin{align}
   \sqrt{2\sinh{(\sqrt{2}|X|})}\cos{Z}=\sigma_z \sqrt{2\sinh{(\sqrt{2}|X|)}-\psi^2},
\end{align}
which follows from the form of the invariant $\psi$. We note several important features of the functions $\psi$ and $ \alpha$. The functions satisfy the parity condition (\ref{parityXYZ}) but are not analytic everywhere. The function $\psi$ is not analytic near $X=0,Z=0,\pm \pi$ while $\alpha$ is not analytic near $X=0$ and $Z=\pm \pi/2$. The lower limit $X_0$ on the integral ensures that $\alpha$ is continuous across $Z=\pm \pi/2$ where $\sigma_z$ is discontinuous. It can be easily verified that (\ref{phi_psi_alpha}) is satisfied when $X\neq0, Z\neq 0$. The integrals appearing in (\ref{Iz},\ref{psi_alpha}) can be evaluated explicitly in terms of incomplete elliptic integral
of the first kind.

\begin{figure}[ht]
\subfigure[\ Contours of $\psi$]{
\includegraphics[width=0.235\textwidth]{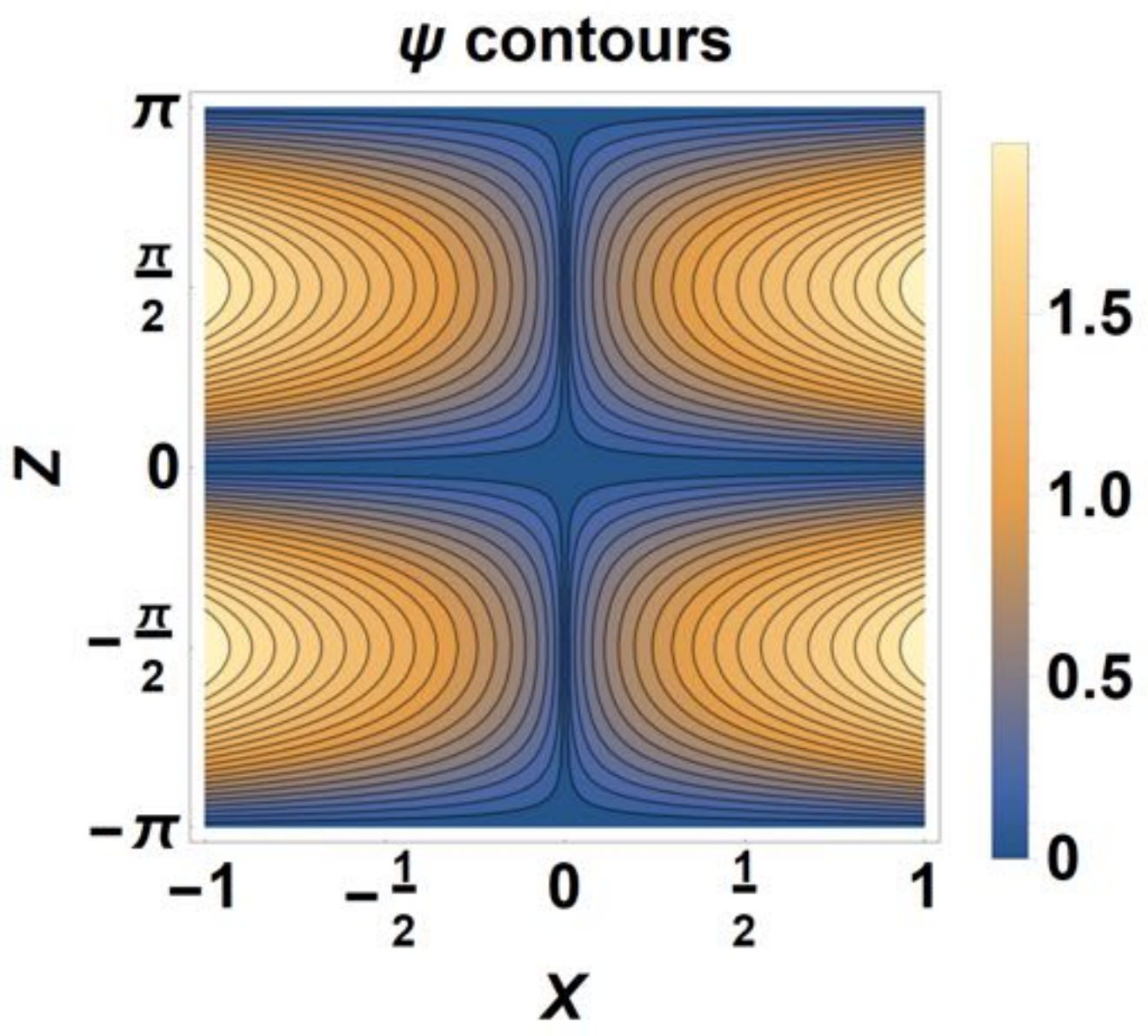}\label{psifig2D}}\hfill
\subfigure[\ Contours of $\alpha$ on $Z=\pi/3$]{
\includegraphics[width=0.235\textwidth]{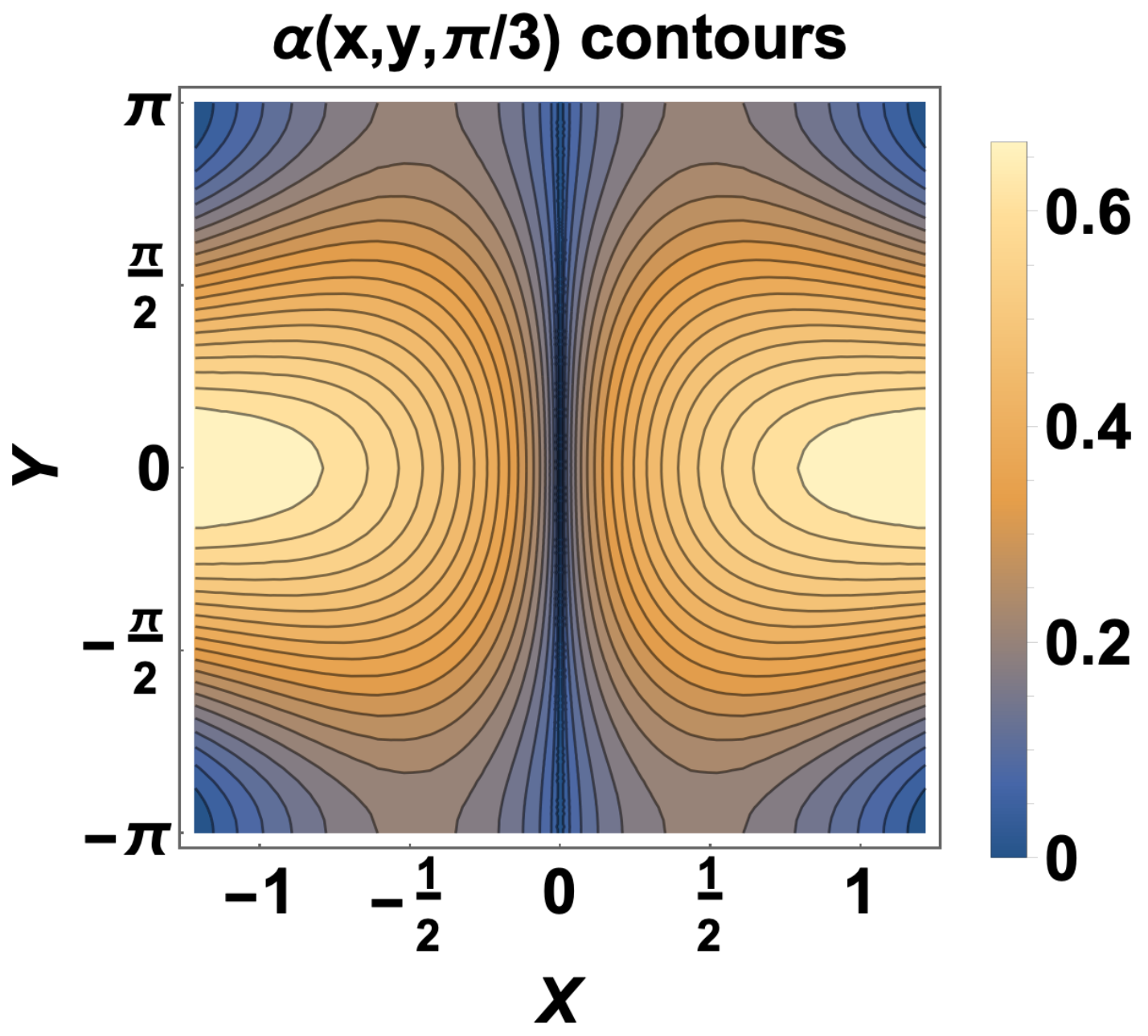}}\\
\subfigure[\ Contours of $\alpha$ on $Z=\pi/2$]{
\includegraphics[width=0.235\textwidth]{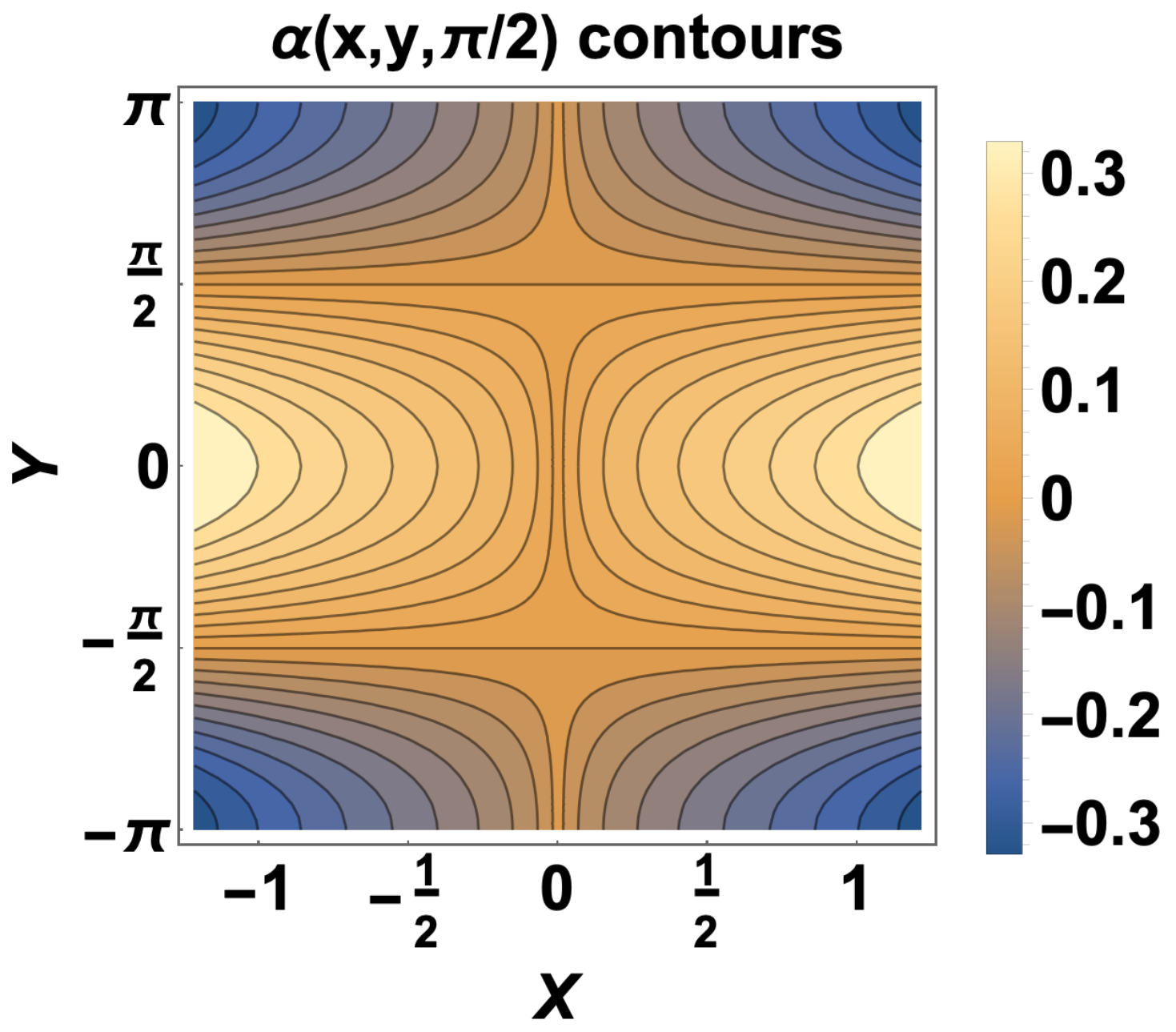}}\hfill
\subfigure[\ Contours of $\alpha$ on $Z=5\pi/6$]{
\includegraphics[width=0.235\textwidth]{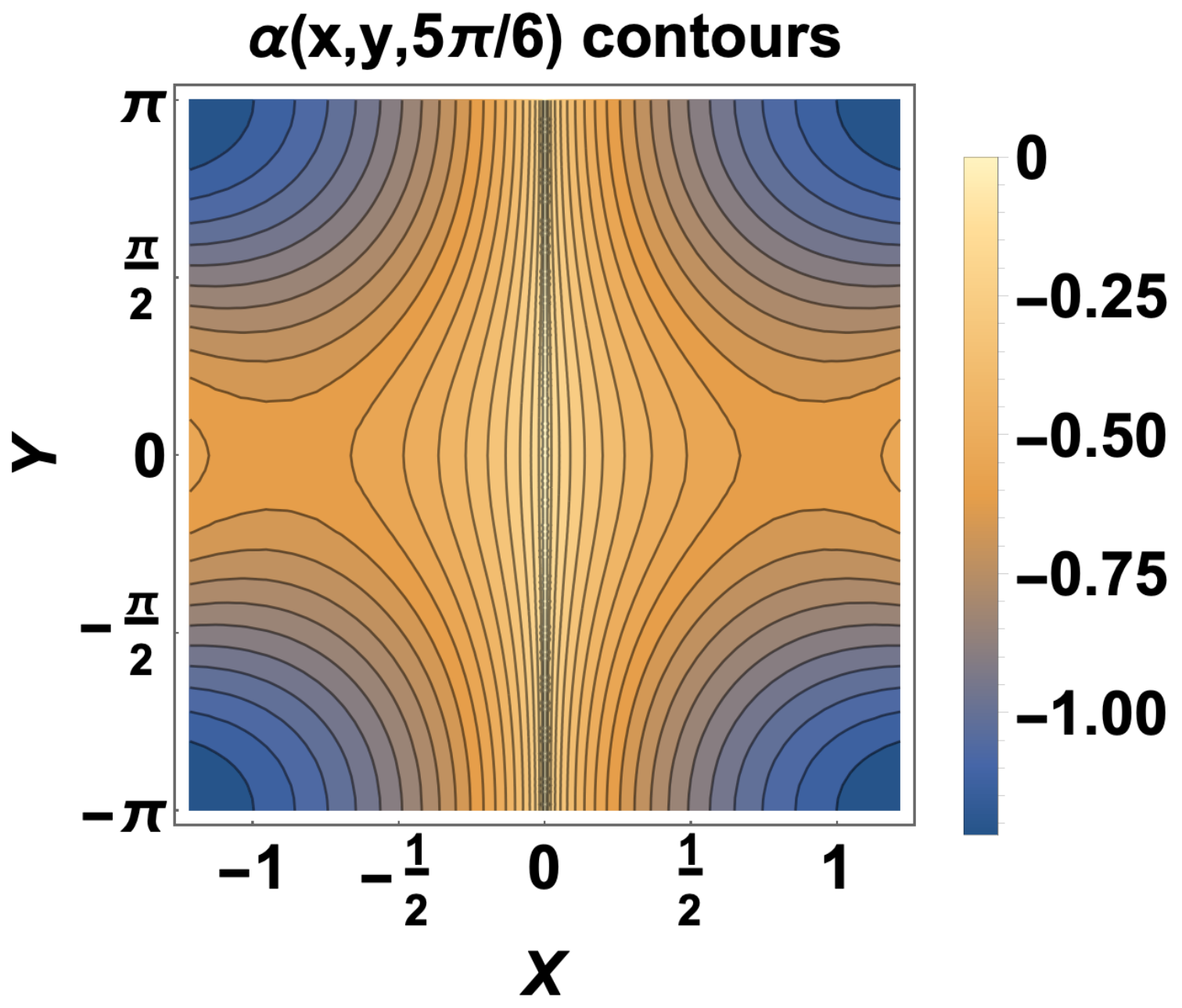}}
\caption{\label{fig:2Dcontours} Contours of the Clebsch variables (Eqns(\ref{modelPhi}),(\ref{phi_psi_alpha})).}
\end{figure}

\begin{figure}
    \centering
    \includegraphics[width=0.45\textwidth]{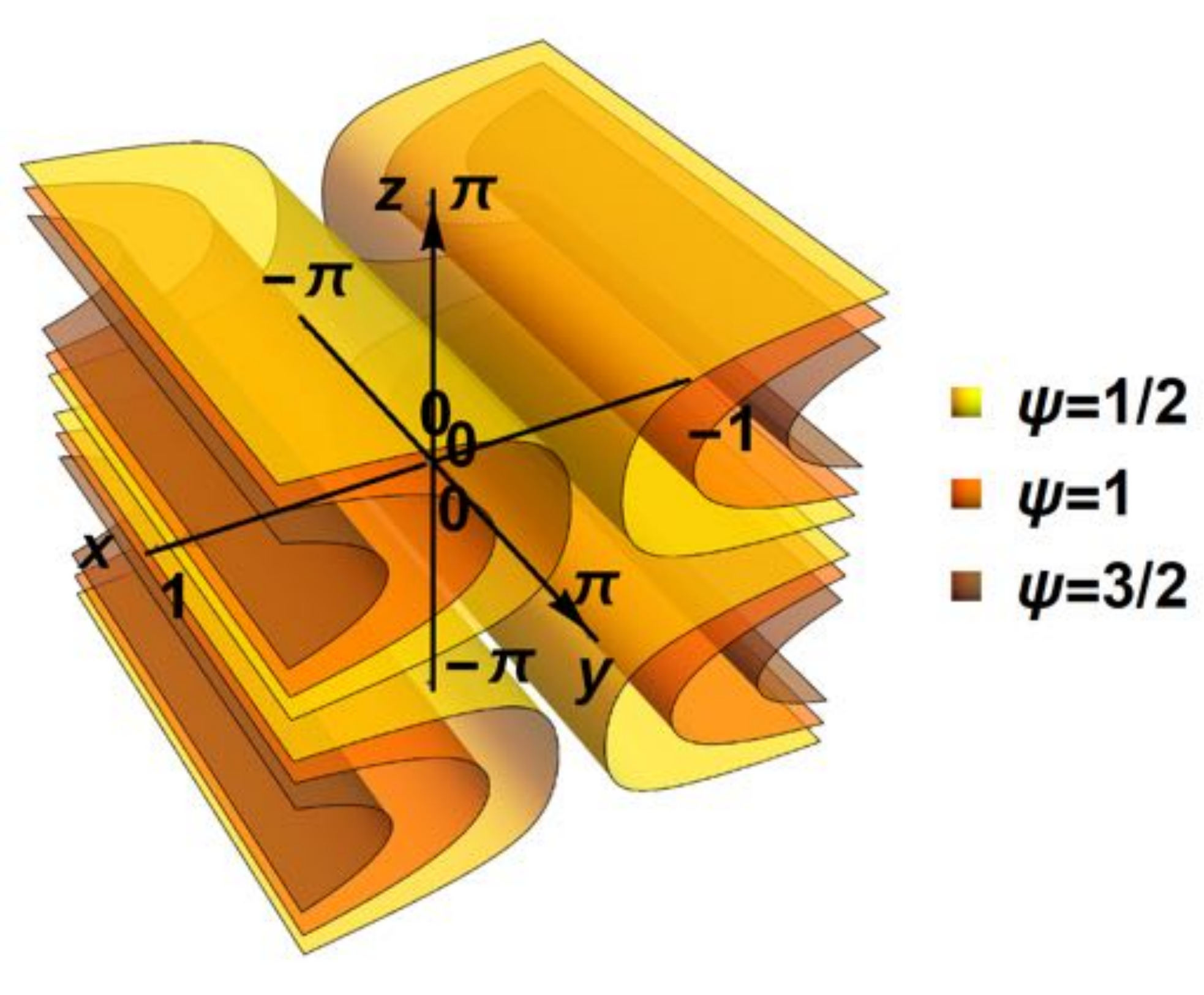}
    \caption{Nested flux surfaces around the X-point at the origin}
    \label{psi3D}
\end{figure}

\begin{figure}
    \centering
    \includegraphics[width=0.45\textwidth]{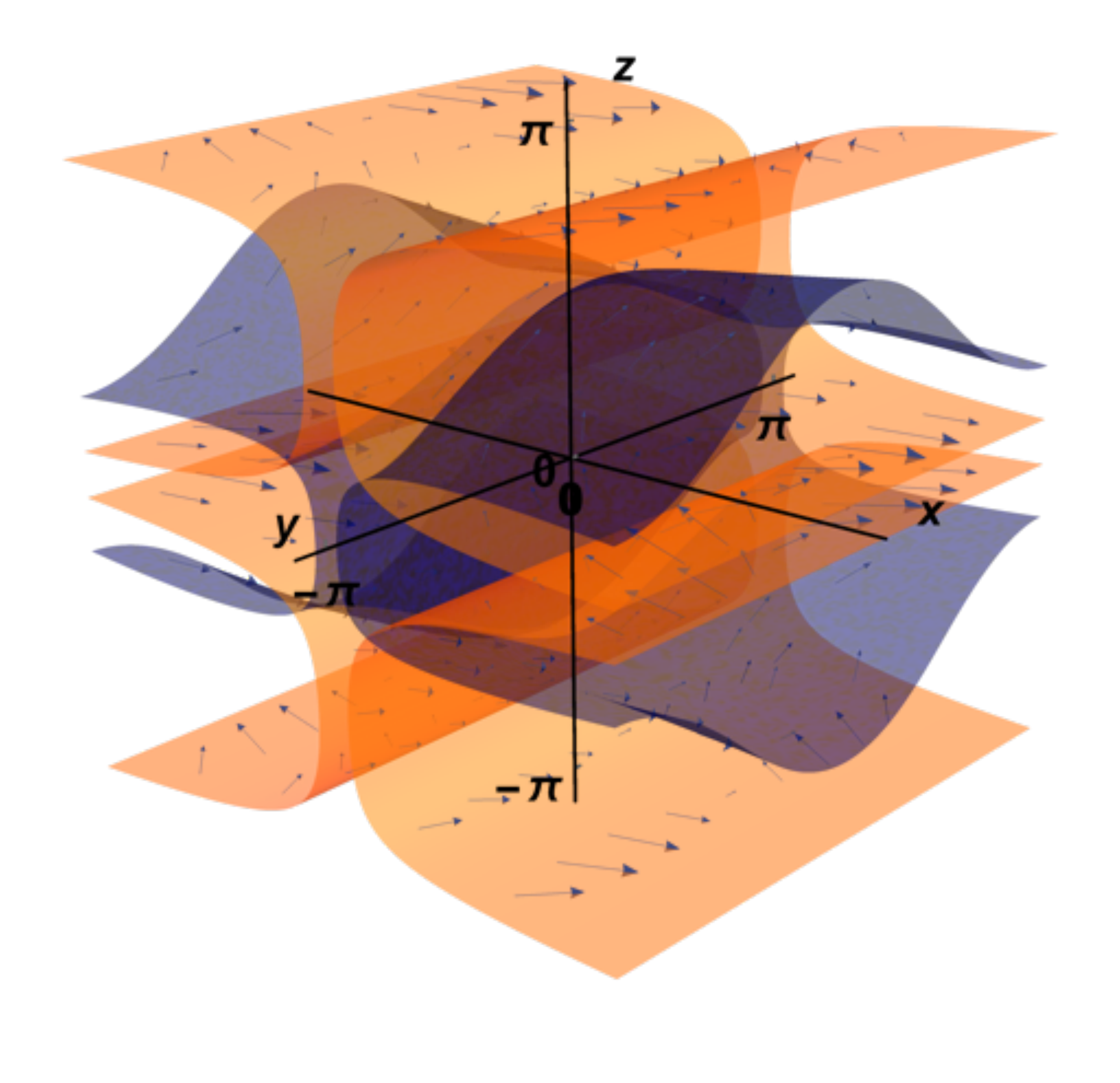}
    \caption{Clebsch surfaces: $\psi$ (orange) and $\alpha$(blue) and the magnetic field lines}
    \label{psi_alpha3D}
\end{figure}

The contours of constant $\psi$, as shown in figure (\ref{psifig2D}) and  (\ref{psi3D}), forms nested surfaces separated by the $X=0$ and $Z=0$ planes and has a X-point at $X=Z=0$. Remarkably, $\psi$ is independent of both the $y$ coordinate and the aspect ratio parameter $\ep$ for this class of fields. The other Clebsch variable, $\alpha$, is manifestly three-dimensional, as shown in figure \ref{psi_alpha3D}.

Near the $X=0$ plane and arbitrary $Y$ and $Z$,
\begin{align}
    \psi &\approx \sqrt{2\sqrt{2}|X|}\sin{|Z|},\label{psi_Ix}\\ 
    \alpha &\approx \frac{\psi}{\sqrt{2}} \:\cI_x = \sqrt{\sqrt{2}|X|}(\ep \cos{Y}+\cos{Z}).\nonumber
\end{align}
Both $\psi$ and $\alpha$ approach zero as $X$ approaches zero but the magnetic field given by $\dl \psi\times \dl \alpha\approx\dl (\psi^2/(2\sqrt{2}))\times \dl \cI_x$ is nonzero.  

Since there is a gauge degree of freedom in choosing the Clebsch potentials such that $\B$ is unchanged, we can choose $\Psi=\psi^2/2$ and $\alphabar=\alpha/\psi$ as new Clebsch potential. From \ref{psi_Ix}, we find that $\Psi$ has second harmonics in (poloidal) angle $Z$ and the hyperbolic dependence on $X$ as expected from near-axis expansions \cite{Solovev1970}. 

On the other hand when we approach the $Z=0$ plane, and $X$ and $Y$ are arbitrary, 
 \begin{align}
    \psi \approx |Z|\sqrt{2\sinh{\sqrt{2}|X|}}, \quad \alpha \approx \cI_Z.
\end{align}
Near the X-point where both $X$ and $Z$ approach zero
\begin{align}
    \psi \approx |Z|\sqrt{2\sqrt{2}|X|}, \quad \alpha \approx \sqrt{\sqrt{2}|X|}(1+\ep \cos{Y}).
\end{align}
In addition to the possible value of explicit examples of doubly periodic vacuum magnetic fields with closed field lines, the results also provide an indirect validation of the formal expansion for vacuum magnetic field flux surface.

\section{The Lortz construction of ideal MHD equilibrium \label{sec:Lortz}}
We next apply the Lortz construction of an ideal MHD toroidal equilibrium as a convergent power series expansion in the parameter beta, starting from a given closed line vacuum magnetic field. Magnetic field line closure is guaranteed order by order in the expansion by the condition that the magnetic fields have reflection symmetry in a given plane. In the present case, the torus is defined by the two ``angles" $Y$ and $Z$, and the system is $2\pi$ periodic in each of the angles. The system will be designed to be symmetric on reflection in the plane $Y=0$, so $B_Y(x=\sqrt{2}X, Y, Z)$ is an even function of $Y$, and the other two components are odd in $Y$. The domain is defined by the relation 
\begin{align}
    x_{-}(Y,Z)< x< x_{+}(Y,Z).
    \label{dom_x}
\end{align}
We discuss the nature of $ x_{-}(y,z)$ and $x_{+}(y,z)$ after further elaboration. We finally obtain equilibrium with boundaries, which are also constant pressure surfaces and with interior pressure maximum or a unique pressure maximum. Neither of these properties holds for the original Lortz analysis. We choose the vacuum field with $a_1=\ep >0, a_n=0, n>1$. We restrict the magnitude of $\ep$ in the subsequent analysis. 

The Lortz analysis is based on the equilibrium representation
\begin{subequations}
\begin{align}
    \dl \cdot \B &=0\\
    \dl \times \B = \J &= \dl \tau \times \dl p\\
    \text{where,}\quad \quad \tau & \equiv \int_0^\ell \frac{d\ell }{B}, \quad  q\equiv\oint \frac{d\ell }{B}, \label{taupq} \\
    p &=p(q). 
\end{align}
\label{Lortz_system}
\end{subequations}
The integral for $\tau$ is along a field line starting on some fixed plane in the domain of interest. We follow the Lortz notation and note that $q$ is not the standard tokamak safety factor. We recall that we start with a given closed line magnetic field, evaluate $\J,q,p$ and then determine a new closed line magnetic field and then iterate. The details are as follows:
\begin{enumerate}
    \item After $n-1$ iterations, we calculate $\tau_n$ and $q_n$ from $\B_{n-1}$. 
\item With $q_n$ known, we update pressure $p_n=p(q_n)$ where, the specific form of pressure is an arbitrary input. 
\item From pressure $p_n$ and $\tau_n$, we update currents using $\J_n=\dl \tau_n \times \dl p_n$.
\item Using $\dl\cdot\B_n=0, \dl \times \B_n = \J_n$, we update $\B_n$ and then iterate.
\end{enumerate}

Provided $p$ is sufficiently small and provided each iterate has reflection symmetry in the plane $Y=0$ the process converges. The additional currents due to finite pressure do not create islands as they are incorporated self-consistently in the pressure profile in each iteration. The periods in $Y$ and $Z$ and hence the rotation transform stay fixed. Closed magnetic field lines, therefore, remain closed even with the addition of pressure and currents.

\subsection{Calculation of q}
It is clear from the preceding discussion that the properties of $\tau$ and especially of $q$ for the initial vacuum field are critical to start the analysis, and we explore the properties of $q$ for our chosen vacuum field. A magnetic field line is given by the solution of the equations
\begin{subequations}
\begin{align}
    \frac{dx}{dY}&=\frac{2\ep}{\cD} \sin{Y}\cos{Z}\sinh{x}\\
    \frac{dZ}{dY}&=-\frac{\ep}{\cD} \sin{Y}\sin{Z}\cosh{x}\\
   \text{where},\quad \cD&= 1+\ep \cos{Y}\cos{Z}\cosh{x},
\end{align}
\label{B_lines}
\end{subequations}
and 
\begin{align}
 x=\sqrt{2}X, \quad q= \int_{-\pi}^{+\pi}\frac{dY}{\cD}.
 \label{qcD}
\end{align}
It is convenient to re-express (\ref{B_lines}) in terms of $\mu=\cos{Y}$. We note that in the range $-\pi<Y<0$, $\mu$ covers the values from -1 to +1 with $\mu$ increasing with $\sin{Y}$ negative, while in the range $0<Y<\pi$, $\mu$ is decreasing with $\sin{Y}$ positive. Thus, we find
\begin{align}
    q&=\int_{-1}^{+1}\frac{d\mu}{\sqrt{1-\mu^2}}\lbr \dfrac{1}{\cD(-\mu)}+\dfrac{1}{\cD(\mu)} \rbr\nonumber\\
    &=2\int_{-1}^{+1}\frac{d\mu}{\sqrt{1-\mu^2}}\dfrac{1}{(1-\ep^2\mu^2\cos^2{Z}\cosh^2{x})}.
    \label{qform}
\end{align}
In the form (\ref{qform}) and using (\ref{B_lines}) it is relatively straightforward to obtain the  needed properties of $q$. 

We study the system (\ref{B_lines},\ref{qcD}) for small $\ep$ and we see that 
\begin{subequations}
\begin{align}
    x&=x_0 +2\mu\ep \cos{Z_0} \sinh{x_0}+(\mu \ep)^2\frac{1}{2}\sinh{(2x_0)}+O(\ep^3)\\
    Z&=Z_0 -\ep \mu\sin{Z_0} \cosh{x_0}+(\mu \ep)^2\frac{1}{2}\sin{(2Z_0)}+O(\ep^3)
\end{align}
\label{xzsmallep}
\end{subequations}
We obtain readily 
\begin{align}
\frac{q}{2\pi}= 1+\frac{1}{2}\ep^2\cos^2{Z_0}\cosh^2{x_0}+O(\ep^4).
\label{q@xpoint}
\end{align}
The critical points of the function $q(x_0,Z_0)$ that occur at $x_0=0,Z_0=0,\pm\pi$ are saddle points. To analyze the behavior of $q$  near $Z_0=\pm \pi/2$ and arbitrary $x_0$, we must expand further. We set $Z_0=\pm\pi/2+\ep\: (\delta Z_0)$ and find
\begin{align}
  \frac{q}{2\pi}=& 1+\ep^4 \cosh^2{x_0}\lbr\frac{3}{8} \cosh^2{x_0}+\frac{1}{2}(\delta Z_0)^2\rbr +O(\ep^6) 
  \label{q@Opoint}
\end{align}
so that $(Z_0=\pm \pi/2,x_0=0)$ is a center. 

\begin{figure}
\subfigure[\ $(2\pi)^{-1}q(x=0)$ as a function of $\cI_x$ (Eq. (\ref{qx0epexp}))]{
\includegraphics[width=0.35\textwidth]{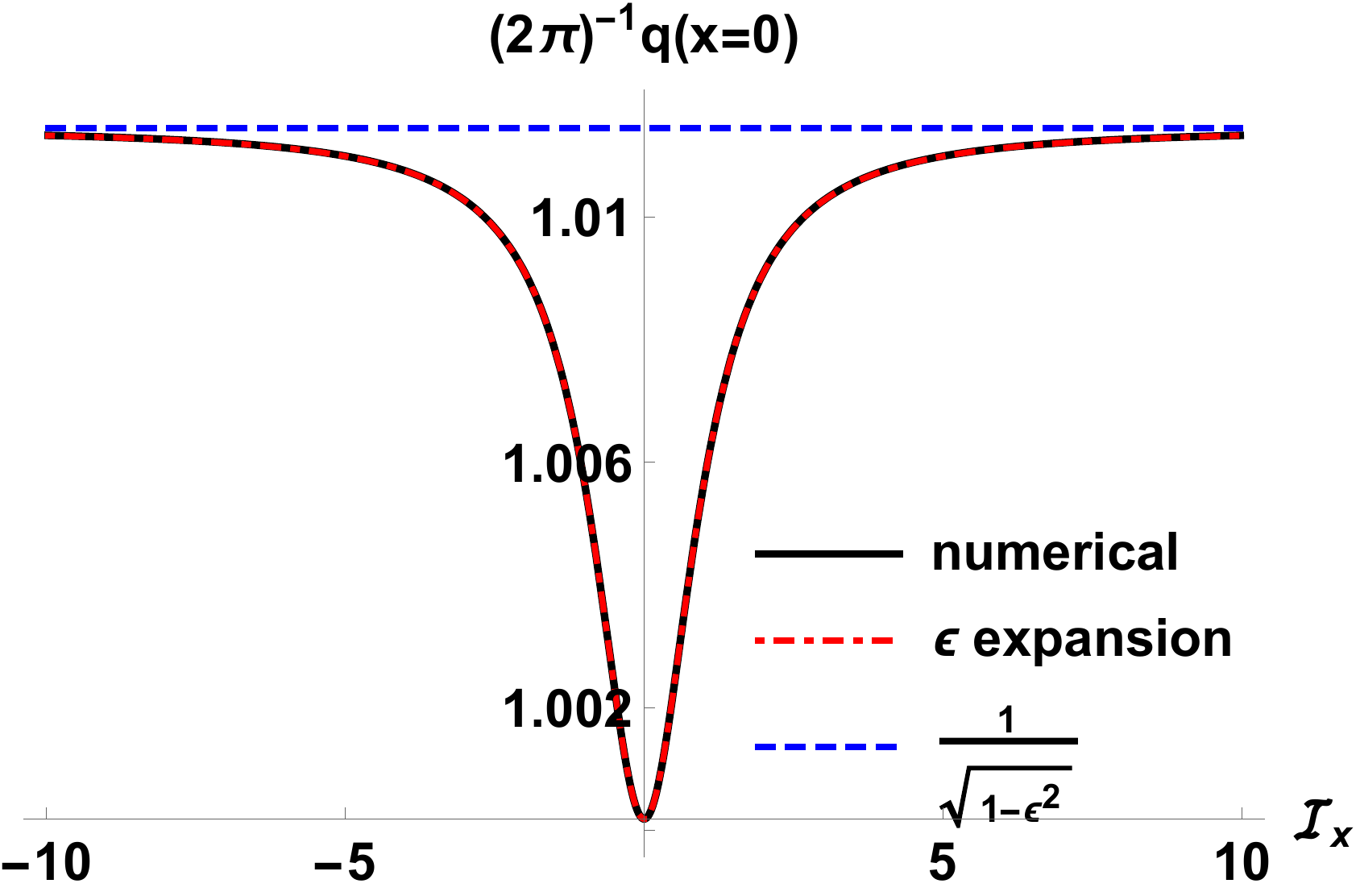}\label{qx0Ix}}\hfill
\subfigure[\ $(2\pi)^{-1}q(x=0)$ as a function of $Z_0$]{
\includegraphics[width=0.35\textwidth]{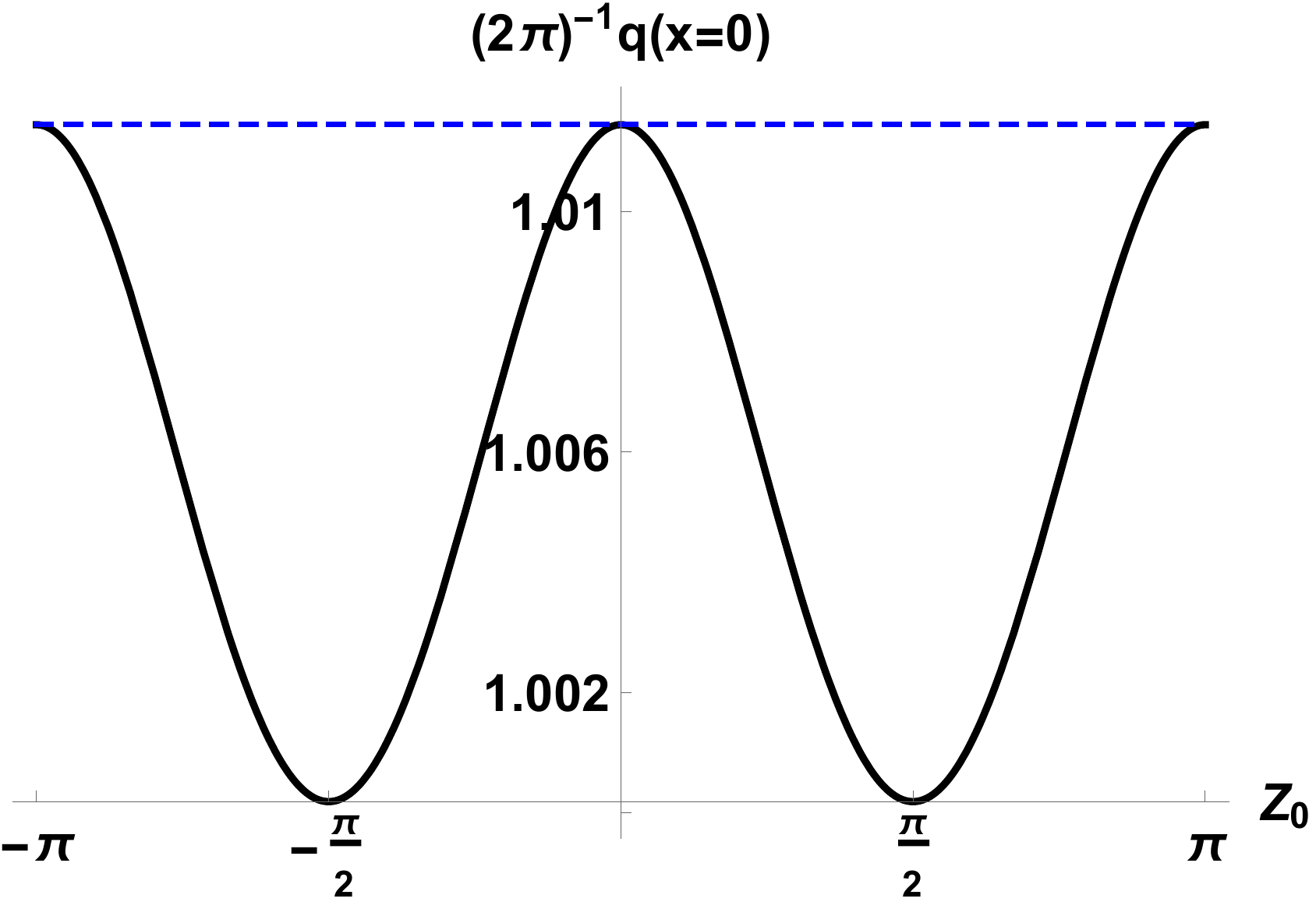}}\label{qx0Zplot}
\caption{\label{qx0plots} Variations of $(2\pi)^{-1}q(x=0)$ with $\cI_x$ and $Z_0$ for $\ep=0.15$}
\end{figure}

We show the variation of $q(x=0)$ with the invariant $\cI_x$ and $Z_0$ in figure \ref{qx0plots}. The function is bounded from above by $1/\sqrt{(1-\ep^2)}$ which is the value of $q$ when both $x$ and $Z$ are zero, i.e., at the X-point. On the other hand, at the centers ($x=0,Z=\pm\pi/2$), $q=2\pi$, as can be seen from (\ref{qform}).

\begin{figure}
    \centering
    \includegraphics[width=0.45\textwidth]{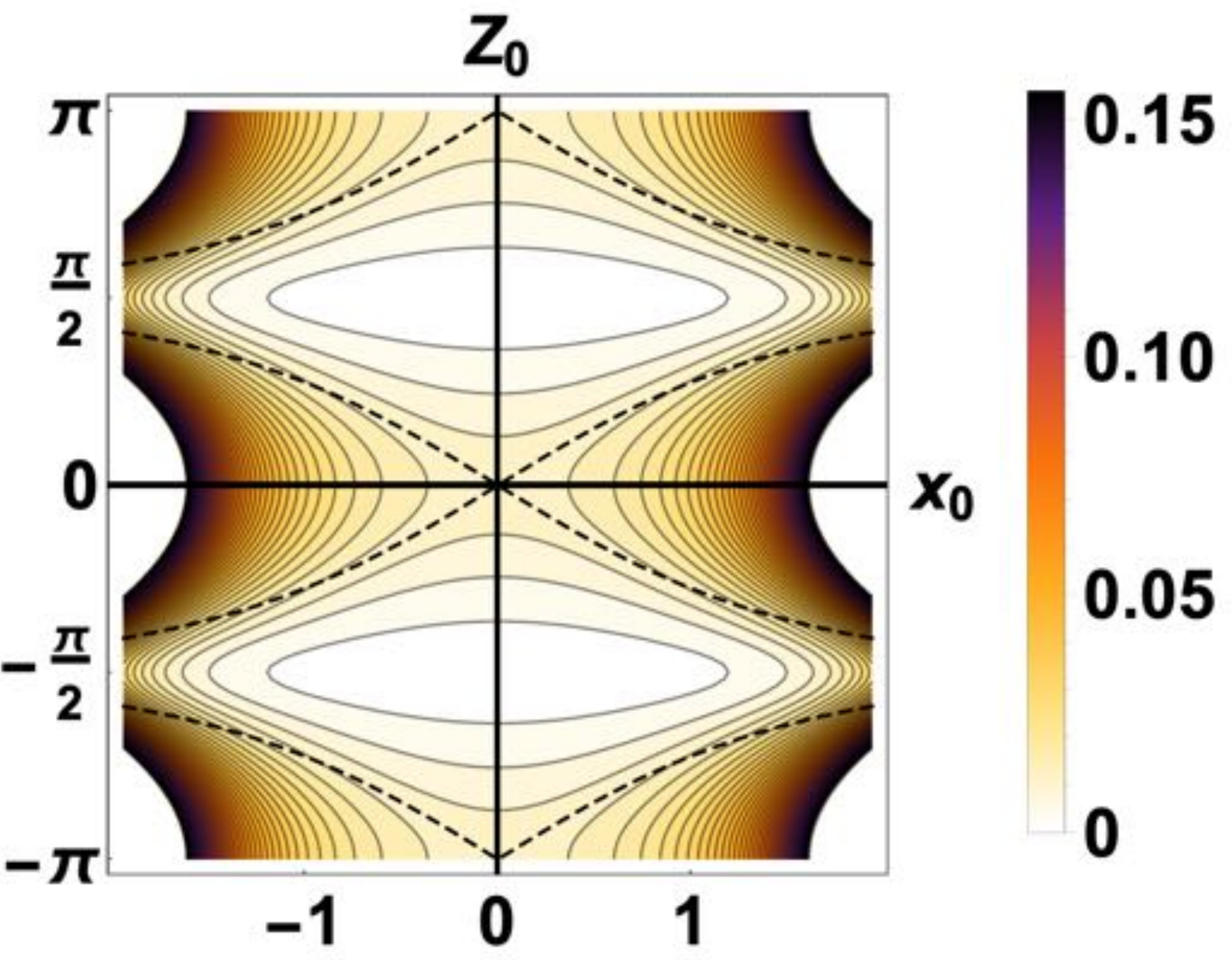}
    \caption{Contours of $\lbr (2\pi)^{-1}q(x_0,Z_0)-1\rbr$ (Eq. (\ref{q4epexp})) is shown when $\ep=0.15$. The approximate separatrix $\cos{Z_0}\cosh{x_0}=\pm 1$ is shown by dashed lines. }
    \label{qcontours}
\end{figure}

The $\ep$ expansion is seen to be in excellent agreement with the numerically obtained value of $q(x=0)$. Figure \ref{qcontours} show the results of the $\ep$ expansion carried to fourth-order. The analytical expressions for various quantities are provided in the appendix. The saddle point at ($x_0=0,Z_0=0,\pm \pi$) and the center at ($x_0=0,Z=\pm \pi$) are clearly visible. We see from (\ref{q@xpoint}) and (\ref{q@Opoint}) that a class of curves $q=$ constant, is approximately given by $\cos{Z_0}\cosh{x_0}=$ constant. In particular, the lines
\begin{align}
\cos^2{Z_0}\cosh^2{x_0}=1
    \label{ep2sept2}
\end{align}
approximate the separatrix joining the saddle point to $O(\ep^2)$.

\begin{figure}
    \centering
    \includegraphics[width=0.45\textwidth]{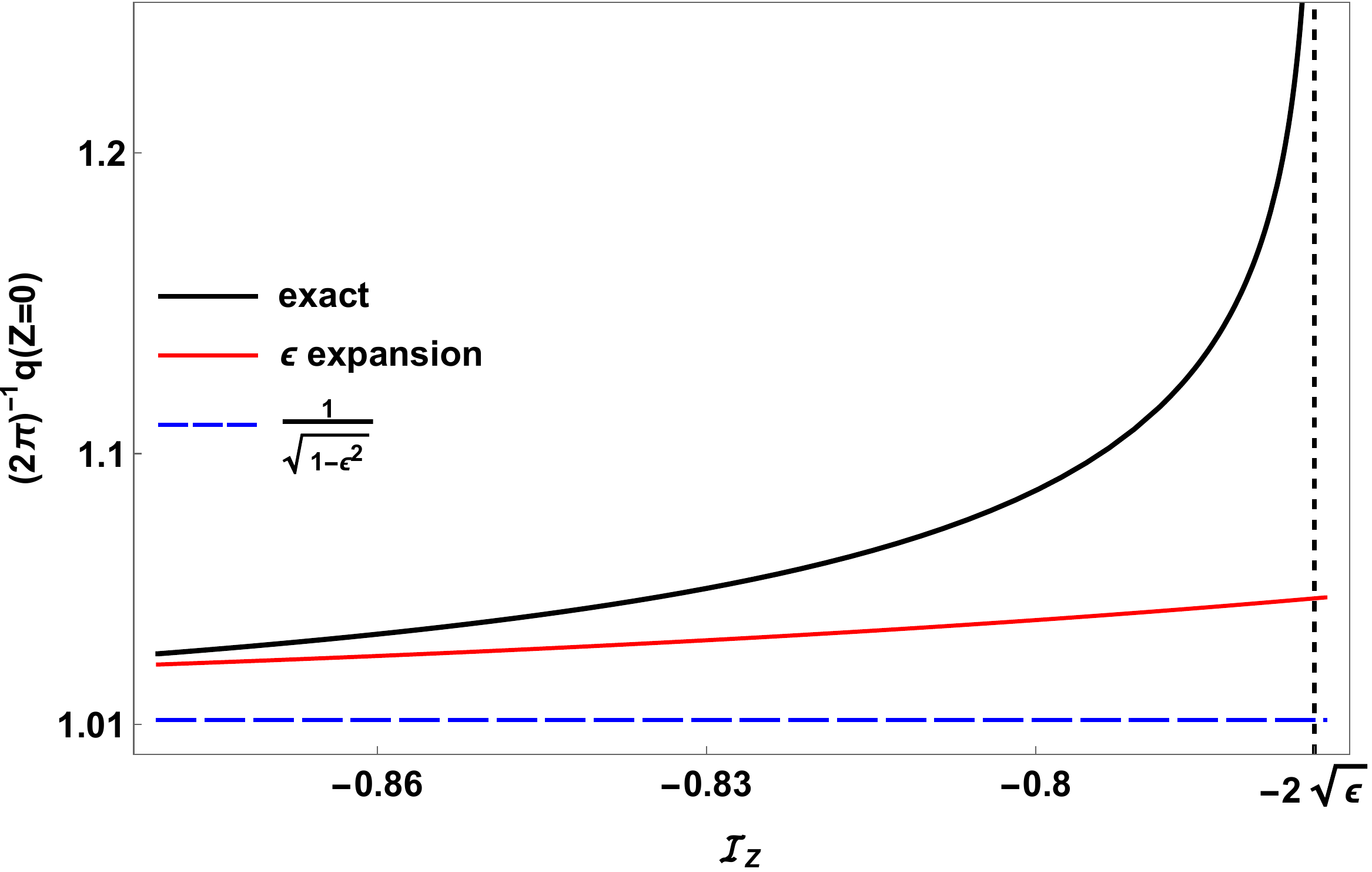}
    \caption{ $\lbr (2\pi)^{-1}q(Z=0)\rbr$ (\ref{qexactz0}) as a function of $\cI_Z$ with $\ep=0.15$. The logarithmic singularity occurs at $\cI_Z=-2\sqrt{\ep}$. }
    \label{figqZ0}
\end{figure}

We observe that for large $|x|$, the contours bunch around a boundary, which is shown in the appendix to be the curve $\cD(\mu=\pm 1)=0$. From the definition of $q$ (\ref{qform}), we see that the perturbation theory can fail for large $x$ as $\cD(\mu)$ approaches zero. Physically, the magnetic field has a turning point near this boundary, and therefore, $q=\oint dl/B$ can become singular. We present a detailed description of the behavior of $q$ near this boundary in the appendix. We show that $q$, in general, has a logarithmic divergence near the boundary $B_Y=0$. As a concrete demonstration of the logarithmic behavior near the boundary, we calculate analytically the value of $q(Z=0)$ for large $x$ without the $\ep$ expansion. Figure \ref{figqZ0} compares the exact expression for $q(Z=0)$ for large $x$ (\ref{qexactz0}) to the result obtained using the $\ep$ expansion. As seen from the figure, near the boundary $\kappa=1$, which occurs when $\cI_Z=-2\sqrt{\ep}$, $q$ diverges logarithmically. The $\kappa=1$ curve is indeed the boundary on which $B_Y$ vanishes.

Finally, we observe that any sufficiently small and sufficiently smooth perturbation of $q$ cannot change the character of $q$. There remain centers in the neighborhood of ($Z_0=\pm \pi/2,x_0=0$). Thus, one might add additional sufficiently small terms from the series (\ref{Phi_series}). Hence, the behavior of $q$ presented here is expected to be modified only slightly when plasma $\beta$ is introduced as a small perturbation to the vacuum magnetic field system in the Lortz iteration scheme.

\subsection{Lortz iteration}

We now restate the critical elements of the Lortz iteration. We solve Eq.(\ref{Lortz_system}) iteratively. The solution of the system is to have reflection symmetry in some planes. That symmetry may be translated into even and odd conditions in the components of $\B$ on the given plane. If the domain is multiply connected, then appropriate periods must be given. Provided the domain boundary and the function $p(q)$ is sufficiently smooth, and provided there is a vacuum field satisfying all the periodicity, periods and symmetry conditions, then for $p(q)$ sufficiently small, a solution of system (\ref{Lortz_system}) exists with appropriate smoothness properties. The iteration starts with $\J=0$, and from the vacuum, $\B$ one calculates $\tau,q,p(q)$ and proceeds. 

We choose the vacuum field discussed in section \ref{sec:model_vacuum}, and we consider the plot for $q$ in the $x,Z$ plane. It is to be noted that in closed field line systems, the Clebsch potential $\psi$ alone does not determine confinement. Since pressure is a function of $q=q(\psi,\alpha)$, confinement is determined by closedness of the $q$ profile. Therefore, the fact that the $\psi$ surfaces shown in figure \ref{psifig2D} have hyperbolic behavior near $X=0$ and are not closed do not mean confinement is not possible.

We shall now describe two different equilibrium constructions. In the first case, we identify right and left boundaries of the domain: each boundary curve runs from $Z=-\pi \:\: \text{to}\:\: Z=+\pi$ and lies outside curves connecting the X points. Each point on the bounding curve corresponds to a magnetic field line, and the totality of these field lines generates the bounding surfaces. Each surface consists of magnetic field lines with the same values of $q$. The field lines are symmetric in $Y$ and about $Y=0$, and thus the domain has the symmetry. We may choose the plane of this curve to correspond to $Y=0$. We may then give $p=p(q)$ for some fixed form of $p(q)$ and carry out the iteration process. The iteration involves small changes in $\B$ or order $\beta$, and thus, the structure of the modified $q=$ constant curves includes at most small variations of the structure of the $q$ surfaces. Therefore, the limit of the Lortz iteration will have a similar $q=$ constant surface. We solve in the fixed domain whose boundary is composed of magnetic field lines, but we cannot guarantee the $q=$ constant on the boundary. However, there will be an adjacent surface inside the domain on which $q$ is constant. Thus, we can construct an equilibrium whose boundaries are pressure surfaces. By an appropriate multiple values choice of $p(q)$, we may ensure that there is a unique pressure maximum. We may lose the properties that the solution is even in $x$ and $Z$. We preserve, however, the periodicity in $Y$ and $Z$.

In the second case, we shall relax the symmetry condition in $Z$. We identify a section of that plane containing the O point and in which field lines through $(x, Y, Z)$ where $(x, Z)$ lies in the plane, can extend from $-\infty$ to $+\infty$ in $x$. Clearly, such domains exist close to $Z=\pm \pi/2$. The bounding curve of the domain is to have many derivatives, and may or may not be a curve of constant $q$. If the three-dimensional domain is to have surfaces of different pressures extending to the boundary, then the chosen curve must be one with constant $q$. We identify the region in the $(x,Z)$ plane with a value $Y_0, -\pi \leq Y_0 \leq \pi$, and we draw field lines through the points $(x,Y_0,Z_0)$. We choose this surface as our boundary of the domain in which to carry out the iteration. Clearly, the domain and the initial vacuum field satisfy the symmetry, periodicity, and period conditions. Thus, we have constructed an equilibrium. We note that although the initial vacuum field had symmetry properties in $Z$, they are lost in this construction.
Further, if we wish to have the boundary be a pressure surface for the equilibrium, then we must require that $p(q)$ be constant for $q$ in some open interval containing the values of $q$ on the initial boundary surface. The Lortz construction cannot guarantee that every field line on the boundary will have the same value of $q$, even if the initial field has the property. Our construction here produces a confined plasma, with either a vacuum region outside of the plasma or equilibrium with $p=$ constant on a boundary surface in the neighborhood of the initial surface chosen. We note in passing that the work of Lortz, on which this material is based, did not address the issue of whether or not the construction produced a confined plasma with a given bounding pressure surface.

We observe from the given vacuum magnetic field that there are many different possible equilibria with the same $p(q)$ but with different boundaries. Our argument of an equilibrium construction applies equally well for vacuum fields of the more general form (\ref{Phi_series}) provided only that the vacuum magnetic field possesses a center in the plot of $q$. As noted, for fields sufficiently close to the one studied in detail, this property holds. 

The Lortz analysis does not require that the given data the boundary of the function $p(q)$ be analytic. While his analysis requires typically $\B$ have Holder continuous second derivatives in all variables, one might extend the work to analytic data in x for which the solutions would also be analytic in $x$, if such exist. The formal expansion of a solution in a power series in $x$ with corresponding data on $x=0$ should converge to this solution. The convergence of the series for other classes of initial data is not credible.

We conclude with an observation concerning another equilibrium representation, which also lends itself to the construction of an iteration scheme :
\begin{align}
    \B =\dl \Phi +\tau \dl p, \quad \J=\dl \tau\times \dl p, \quad \dl\cdot \B=0
    \label{gradBform}
\end{align}
where $\tau,p$ and $q$ are again given by (\ref{taupq}). Suppose $\B$ has $N$ Holder continuous derivatives. Then $\tau$ and thus $p(q)$ would have in general $N-1$ Holder continuous derivatives. Hence (\ref{gradBform}) would require that $\Phi$ has $N-1$ Holder continuous derivatives but the sum on the right-hand side of (\ref{gradBform}) would require $N$ such derivatives. Such a structure is possible but highly peculiar. Perhaps the iteration based on (\ref{gradBform}) can recover only analytic solutions. 

\section{Conclusion}
Modern stellarators are designed through the use of sophisticated numerical optimization tools. However, minimal analytical insights are obtained in such a process. Such insights, although useful to designers and physicists \cite{landreman_Sengupta2018direct,landreman_Sengupta_Plunck2019direct}, are hard to obtain because of the challenges due to the three-dimensional geometry.  Our main focus here has been to obtain analytical results that can help shape our intuitions and to show that MHD equilibrium with smooth pressure and closed field lines can be constructed following Lortz's construction. We have considered a simplified model of a stellarator:  a toroidal shell with Euclidean metric and periodic boundary conditions in the two angles $Y$ and $Z$. Any effects of toroidal curvature are absent from our analysis. In a toroidal geometry, double-periodicity imposes critical constraints on the magnetic field structure, and our simple model allows us to analyze this feature fully.

In this work, we have constructed vacuum magnetic fields in a topological torus where all the magnetic field lines close on themselves. We have shown that following Lortz, we can obtain an MHD equilibrium with closed field lines as well. Our model is relevant to present experiments like W7-X in the  ``high-\iotabar" configuration, which can access low-shear near rational transform. We have shown that for a class of these vacuum fields, the Clebsch coordinates $\psi$ and $\alpha$ can be constructed explicitly. Analysis of the $\psi$ surfaces shows the existence of an X-point. The behavior of the fields and the Clebsch potentials near the singular line $X=0, Z=0$, has been discussed. The exact solutions obtained in this work, therefore, might be of interest in studying charged particle motion near an X-point, or the separatrix. 

We have discussed an extension of Lortz's iterative construction of ideal MHD equilibrium, starting with a vacuum field with zero rotation transform and adding plasma beta as a perturbation in a topological torus. Our construction allows the rotation transform to be any arbitrary rational number. We have provided analytical expressions for the various quantities involved in the Lortz construction, e.g., the Clebsch variables and the quantity $q=\oint dl/B$. In the future, we shall investigate the possibility of constructing low magnetic shear ideal MHD equilibrium through a Lortz-like iterative scheme in a topological torus, starting with the closed line vacuum magnetic field and adding both the plasma beta and the magnetic shear are treated as perturbations. 

\begin{acknowledgments}
This research was funded by the US DOE Grant No. DEFG02-86ER53223.
\end{acknowledgments}

\appendix*

\section{Details of the calculation of $q$}
We present here some details of the calculation of $q$. We shall obtain asymptotic expressions of $q$ for various regions of interest.
\subsection{Evaluation of $q$ at $x=0$}
The function $q(x=0)$ can be evaluated exactly from (\ref{qform}) with the help of (\ref{Ix}). However, the analytical expression without the small $\ep$ expansion is rather cumbersome. We present here only the $\ep$ expansion result. When $x=0$, using (\ref{Ix}) and (\ref{qform}) we obtain
\begin{align}
q(x=0)= 2\int_{-1}^{+1}\frac{d\mu}{\sqrt{1-\mu^2}}\dfrac{1}{(1-\ep^2\mu^2\cos^2{Z})}\\
\text{with}\quad Z=\cot^{-1}(\cI_x)+\sin^{-1}\lbr\dfrac{\ep \mu}{\sqrt{1+\cI^2_x}}\rbr. \nonumber
\end{align}
Expanding in $\ep$, we get 
\begin{align}
\dfrac{q(x=0)}{2\pi}-1=\frac{\cI_x^2 \epsilon ^2}{2 \left(\cI_x^2+1\right)}+\frac{3 \left(\cI_x^4-\cI_x^2+1\right) \epsilon ^4}{8 \left(\cI_x^2+1\right)^2}\label{qx0epexp}\\
Z=Z_0+\ep \mu \sin(Z_0)+\dfrac{1}{6}(\ep \mu \sin(Z_0))^3+O(\ep^5)\\
 \text{where}\quad Z_0=\cot^{-1}(\cI_x). \nonumber
\end{align}
\subsection{Evaluation of $q$ using the $\ep$ expansion}
We note that only second order corrections are needed from $x$ and $Z$ to calculate $q$ to $O(\ep^4)$. Using straightforward $\ep$ expansion, we obtain
\begin{align}
    \frac{q}{2\pi}-1&= \frac{1}{2}\cos^2{Z_0}\ep^2\cosh^2{x_0}+ \frac{3}{512} \ep^4 \left( 9 - 28 \cosh(2 x_0) \right. \nonumber\\
  +&  \cos(4 Z_0) ( 23 -4 \cosh(2 x_0) + 5 \cosh(4 x_0))\nonumber\\
  +& 32 \left.\cos(2 Z_0) ( 3 + 7 \cosh(2 x_0)) \sinh^2 x_0 \right) \nonumber\\
  +& 59 \cosh(4 x_0)) +O(\ep^6).
  \label{q4epexp}
\end{align}
\subsection{Beyond the $\ep$ expansion: behavior of $q$ near $B_Y=0$: }
\begin{figure}
    \centering
    \includegraphics[width=0.4\textwidth]{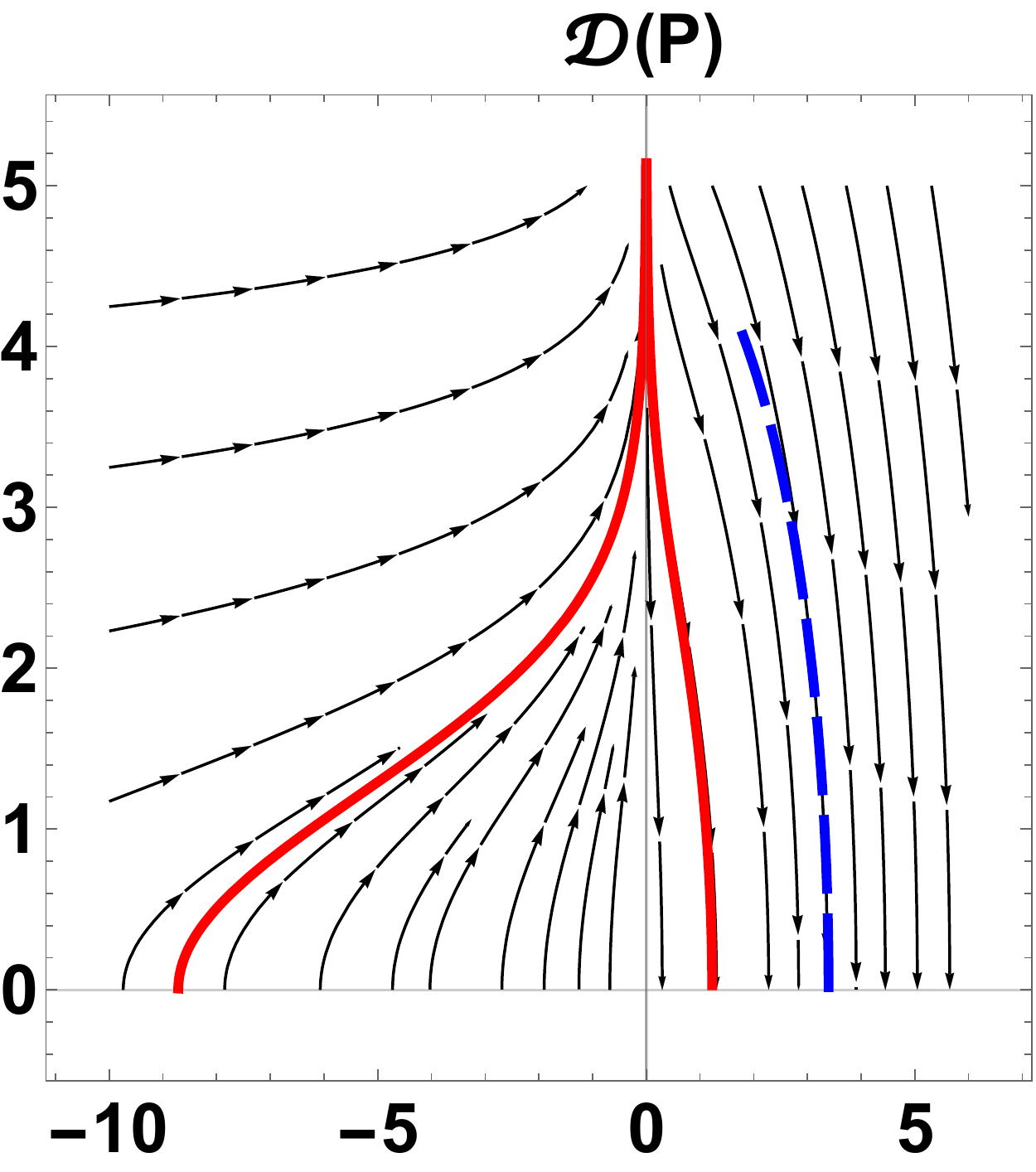}
    \caption{Phase diagram of (\ref{cDPdotted}) near $\cD(P)\approx 0$. In thick red we show a typical solution (\ref{cDP}), and in dashed blue we show a typical approximate solution (\ref{cDPquad}) }
    \label{DmuP}
\end{figure}

We have noted earlier that the domain of interest is in a region where $B_Y$ has no turning points. However, for sufficiently large $x$, $B_Y$ can approach zero when $\cD(\mu)\approx 0 $, which occurs near the curve
\begin{align}
 \cos Z \cosh x \approx -\frac{1}{\ep \mu}.
\end{align}
Since, both $\mu$ and $ \cos Z$ are $O(1)$ quantities, $\ep\mu \ll 1$ implies $\cosh x\gg 1$. 
It is convenient to recast (\ref{B_lines}) in the form
\begin{align}
-\dfrac{1}{2}\dfrac{d \log{\sinh x}}{d(\log{\ep \mu})}=\ep \mu \cos Z \cosh x \nonumber\\
\dfrac{d \log{\sin Z}}{d(\log{\ep \mu})}=\ep \mu \cos Z \cosh x .
    \label{B_line_logs}
\end{align}
Using 
\begin{align}
    \ep \mu \cos Z \cosh x=1-\dfrac{1}{\cD} \nonumber\\
    \dfrac{d \log{\cosh x}}{d(\log{\ep \mu})}=\tanh^2{x}\dfrac{d \log{\sinh x}}{d(\log{\ep \mu})} \label{use_ident}\\
    \dfrac{d \log{\cos Z}}{d(\log{\ep \mu})}=-\tan^2{Z}\dfrac{d \log{\sin Z}}{d(\log{\ep \mu})}, \nonumber
\end{align}
we obtain the exact result
\begin{align}
    \dfrac{d\cD}{d \log{\ep \mu}}= (\cD-1)\lbr 1-(\tan^2{Z}+2\tanh^2{x})\lbr \dfrac{\cD-1}{\cD}\rbr\rbr.
    \label{cDdotexact}
\end{align}
Near $\cD \approx 0$, $x\gg 1$, we have $\tanh x \approx 1$ and therefore,
\begin{align}
    \psi^2=2 \sinh x \sin^2 Z \approx 2\cosh{x} \sin^2 Z \nonumber \\ =\dfrac{(\cD-1)}{\ep \mu}\dfrac{\tan^2 Z}{\sqrt{1+\tan^2 Z}}.
    \label{replacepsi}
\end{align}
Since the $q$ integral treats $\psi$ and $\alpha$ as constants, (\ref{replacepsi}) allows us to solve for $Z$ in terms of $\psi$.
Simplifying (\ref{cDdotexact}) near $\cD\approx 0$ and using (\ref{replacepsi}) we get
\begin{align}
   \dfrac{P}{2} \dfrac{d\cD^2}{dP}\approx -\lbr 2+P(P+\sqrt{P^2+1})\rbr   \label{cDPdotted}\\
   \text{where,} \quad P=\dfrac{\ep \mu \psi^2}{4}. \nonumber
\end{align}
Equation (\ref{cDPdotted}) can be readily integrated and we get
\begin{align}
\cD(P)=\sqrt{\cD_0^2-P(P+\sqrt{P^2+1})-\sinh^{-1}P-4 \log{|P|}},
    \label{cDP}
\end{align}
where, $\cD_0$ is an integration constant. Figure (\ref{DmuP}) shows the phase diagram of (\ref{cDPdotted}) and a typical solution. We note that there are two values of $P$ for which $\cD(P)$ vanishes. They are not symmetric because (\ref{cDPdotted}) is not symmetric about $P=0$. The positive $P$ solution is the smaller root. The smaller positive root dominates when we calculate $(1/\cD(\mu)+1/\cD(-\mu))$.

For large $P$, the $\sinh^{-1}$ and the logarithmic terms are both small, and it can be seen that 
\begin{align}
    \cD(P)\approx \sqrt{\cD^2_0-2P^2}.
    \label{cDPquad}
\end{align}
Figure (\ref{DmuP}) shows that the approximation captures the $\cD(P)\approx 0$ behavior quite well. A slightly more sophisticated analysis involving the logarithmic term can be done, but essentially, the roots of $\cD(P)$ are determined mostly by the quadratic powers of $P$ in $\cD(P)$. Therefore,

\begin{align}
    q=\int_{-1}^{+1}\frac{d\mu}{\sqrt{1-\mu^2}}\lbr \dfrac{1}{\cD(-\mu)}+\dfrac{1}{\cD(\mu)} \rbr \nonumber\\
    \approx 2\int_{-1}^{+1}\frac{d\mu}{\sqrt{1-\mu^2}\sqrt{c^2-\mu^2}}=\dfrac{2}{c}\text{K}(1/c^2),
\end{align}
where $c^2=(\cD^2_0/2)/(\ep \psi^2/4)$ and K is the complete elliptic integral of the first kind. The above analysis shows that 
$q$ is well-defined except at $c=1$, where it has a logarithmic singularity. The point $c=1$ corresponds to $\cD(\mu)=0$ at $\mu=\pm 1$. 

\begin{figure}
    \centering
    \includegraphics[width=0.5\textwidth]{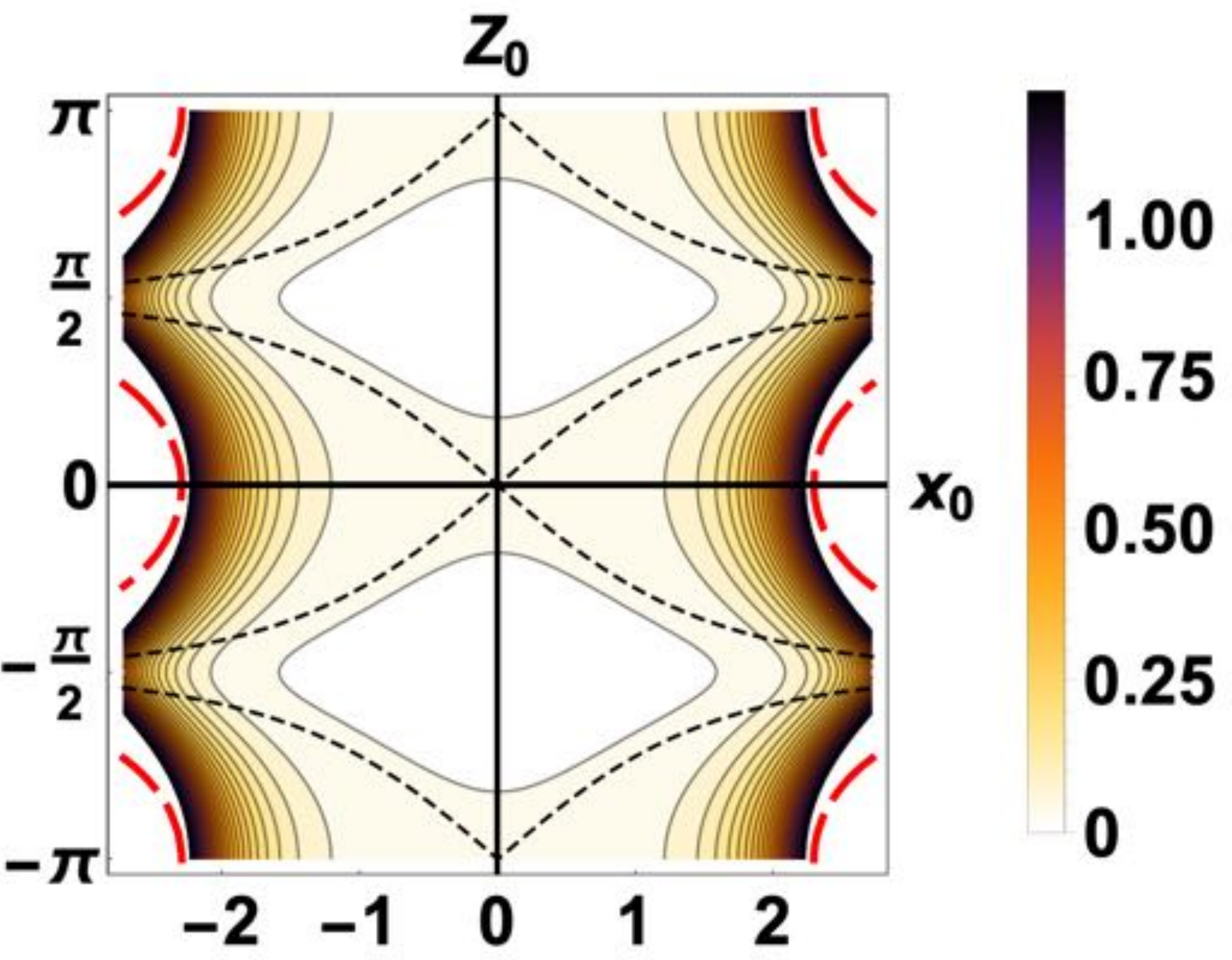}
     \caption{Contours of $\lbr (2\pi)^{-1}q(x_0,Z_0)-1\rbr$ (Eq. (\ref{q4epexp})) for $\ep=0.15$ is shown along with the approximate separatrix (dashed black lines). The boundary $\cD(\pm 1)=0$ is shown by thick dashed red lines }    \label{qsepDmuP}
\end{figure}

We show the curves of $\cD(\pm 1)=0$ along with the result from the $\ep$ expansion in figure \ref{qsepDmuP}. The curves $\cD(\pm 1)=0$ extend to $x=\pm \infty$ at $Z=\pm \pi/2$. Although the $\ep$ expansion fails near the $\cD(\pm 1)=0$ boundary, it correctly predicts that $q$ should increase as we approach the boundary, as can be seen from the bunching of the contours near the boundary.

\subsection{Evaluation of $q$ at $Z=0$}
Independent verification of the logarithmic behavior can be obtained by calculating $q$ for $Z=0$. We restrict ourselves only to relatively large values of $x$ so that $\sinh x\approx (1/2)e^x$. We then find from (\ref{Iz}) that
\begin{align}
    \ep \mu=2\lambda\lbr \lambda+\dfrac{\cI_Z}{\sqrt{2}} \rbr \quad \text{where,} \quad \lambda = e^{-x/2}.
    \label{lambdadef}
\end{align}
Since $\lambda\geq 0$, and $-1\leq \mu \leq 1$, $\cI_Z$ must be negative. In terms of $\lambda$, $q$ is given by
\begin{align}
q=-\frac{4\sqrt{2}}{\ep I_Z}\int_{\lambda_{-}}^{\lambda_{+}} \dfrac{\lambda^2\: d\lambda}{\sqrt{1-\frac{4 \lambda^2}{\ep^2}\lbr \lambda+\frac{\cI_Z}{\sqrt{2}} \rbr^2}}
    \label{qinlambda}\\
\text{where,} \quad \lambda_{\pm}=-(\cI_Z+\sqrt{\cI_Z^2\pm 4\ep})/\sqrt{8}.    \nonumber
\end{align}
The integral can be calculated exactly and we get
\begin{align}
q(Z=0)=2\pi-\frac{\sqrt{2}\Lambda^+}{\cI_Z }\left( (\cI_Z^2-a^{-}a^{+})\text{E}(\phi,\kappa)
+\right.\nonumber \\
\left.\left(a^{-}a^{+}-2 \cI_Z a^{-}+\cI_Z^2 \right)\text{F}(\phi,\kappa)  +4 \cI_z a^{-}\Pi(\Lambda^+,\phi,\kappa) \right) \label{qexactz0}\\
\text{where,} \quad a^{\pm}=\sqrt{\cI_Z^2\pm 4 \epsilon },\:\Lambda^\pm=\frac{1}{4\ep}\sqrt{\cI_Z^2\pm a^{-}a^{+}}\nonumber \\
\phi=\sin^{-1}\lbr \frac{\Lambda^-}{\Lambda^+}\rbr,\: \kappa = \frac{2\:\ep \Lambda^+}{\left(a^{+}-a^{-}\right)^2}.
\end{align}
E,F and $\Pi$ are standard elliptic integrals. Near $\kappa=1$ or equivalently, $\cI_Z=-2\sqrt{\ep}$, q diverges since the elliptic integrals have logarithmic singularities. 

To make connections with the $\ep$ expansion, we observe from (\ref{Iz}) that to lowest order in $\ep$, 
\begin{align}
 x_0 \approx-2\ln{\lbr\frac{-\cI_Z}{\sqrt{2}}\rbr}.
\label{largeXIZ}
\end{align}

It is clear that the perturbation expansion is a suspect near $x_0\approx \ln{(2\ep)}$. For $Z=0$ and large $x$,
\begin{align}
    \cD(\mu)\approx 1+2\ep\, \mu\, e^x.
\end{align}
Hence, the logarithmic singularity appears precisely at the $\cD=0$ boundary.

\subsection{Magnetic field circulation, enclosed currents and $q$}
We now focus on the currents required to maintain the closed magnetic field configurations. It might not be straightforward to relate the magnetic field circulation to the enclosed current via the Stokes theorem if the closed field lines do not bound an orientable area. However, there exists a close relationship between the magnetic field circulation and the function $q= \oint dl/B$, which we will now discuss.

The scalar potential $\Phi$, and the current potential $\tau$ (given by \ref{Lortz_system} ,\ref{gradBform}), and their gradients are in general multi-valued functions. Therefore, for both $\B$ and $\J$ to be single-valued, the two potentials need to satisfy certain additional constraints. Following Grad \citep{grad1971plasma}, we find that $\Phi$ and $\tau$ must be of the form
\begin{align}
    \Phi &= F(q)y + G(q)z+ \tilde{\Phi}(q,y,z)\nonumber\\
   p'(q) \tau &= -F'(q)y-G'(q)z +\tilde{\tau}(q,y,z),
    \label{Phitauforms}
\end{align}
where $\tilde{\Phi},\tilde{\tau}$ are periodic in the angles $y$ and $z$. For vacuum fields, $p'(q)$ is zero, therefore, $F,G$ are constants. The above forms are also valid when the field lines are ergodic \cite{weitzner2014ideal}. The function $q$ is then replaced by the flux-surface label $\psi$.

The magnetic circulation is given by
\begin{align}
    \oint \B \cdot \bm{dl}=\oint (\dl \Phi + \tau p'(q)\dl q) \cdot \bm{dl}= [\Phi],
    \label{Bdotdl}
\end{align}
where $[\Phi]$ denotes the net change in the multi-valued function $\Phi$ after the loop integral. We note that the pressure term does not contribute because $\bm{dl}$ is along a closed field line which is perpendicular to $\dl q$. If the field line closes on itself after $n$ poloidal ($y$) and $m$ toroidal ($z$) circuit, we find from (\ref{Phitauforms}) that
\begin{align}
    [\Phi]= 2\pi (n F(q)+m G(q))=[\Phi](q).
    \label{Phibox}
\end{align}
The net circulation of the closed magnetic field, and possibly the net current enclosed by the loop, are therefore, functions of $q$.

For vacuum fields, $$\oint \B \cdot \bm{dl}=\text{constant},$$ as $F, G$ are constants. Since the current is zero in the interior of the domain, there must be currents in the exterior that maintain the vacuum region in the interior. As an example, consider the case of an axisymmetric toroidal shell with a purely toroidal vacuum field in the interior due to a poloidal current distribution on the boundary. Through proper toroidal and poloidal cuts, we can map the toroidal shell onto a Cartesian flat-torus, with $y$ in the toroidal direction, $z$ in the poloidal direction, and $x_a<x<x_b$ labeling the various shells. The toroidal field lines will map onto straight field lines in the interior, pointing in the $y$ direction. The poloidal currents will map onto the outer boundary $x=x_b$, pointing in the $z$ direction. This example is analogous to our case when $\ep=0$. Since many different configurations of currents can produce the same magnetic field, we shall focus only on sheet currents on the boundary. We choose a constant $q=q_b$ surface as a boundary such that $B_y>0$ at all interior points. Since $\BD q=0$, the interior magnetic fields are tangential to the boundary. We prescribe the magnetic field to be the gradient of the scalar potential inside and zero outside of the $q_b$ surface, i.e.
\begin{align}
    \B=\dl \Phi\: \Theta(q_b-q), \quad \J= \dl q\times \dl \Phi\: \delta(q_b-q).
\end{align}
Here, $\Theta$ and $\delta$ functions denote Heaviside theta and Dirac delta functions respectively. 

For MHD equilibrium, there must be currents in the interior to maintain force balance. From the definition of $\tau$ and $q$ (Eq. \ref{Lortz_system}), we have
\begin{align}
     q=\oint \frac{dl}{B} =\oint \dl \tau \cdot \bm{dl}= [\tau].
\end{align}
From (\ref{Phibox}) and (\ref{Phitauforms}), we obtain
\begin{align}
   q p'(q)=-2\pi(n F'(q)+m G'(q))= -[\Phi]'.
    \label{qconstrain}
\end{align}
Therefore, for given function $q$ and the pressure profile $p(q)$, the circulation for MHD equilibrium magnetic fields is given by
\begin{align}
    \oint \B \cdot \bm{dl}= -\int q p'(q) dq +\text{constant.}
\end{align}

\bibliography{plasmalit} 

\begin{thebibliography}{27}%
\makeatletter
\providecommand \@ifxundefined [1]{%
 \@ifx{#1\undefined}
}%
\providecommand \@ifnum [1]{%
 \ifnum #1\expandafter \@firstoftwo
 \else \expandafter \@secondoftwo
 \fi
}%
\providecommand \@ifx [1]{%
 \ifx #1\expandafter \@firstoftwo
 \else \expandafter \@secondoftwo
 \fi
}%
\providecommand \natexlab [1]{#1}%
\providecommand \enquote  [1]{``#1''}%
\providecommand \bibnamefont  [1]{#1}%
\providecommand \bibfnamefont [1]{#1}%
\providecommand \citenamefont [1]{#1}%
\providecommand \href@noop [0]{\@secondoftwo}%
\providecommand \href [0]{\begingroup \@sanitize@url \@href}%
\providecommand \@href[1]{\@@startlink{#1}\@@href}%
\providecommand \@@href[1]{\endgroup#1\@@endlink}%
\providecommand \@sanitize@url [0]{\catcode `\\12\catcode `\$12\catcode
  `\&12\catcode `\#12\catcode `\^12\catcode `\_12\catcode `\%12\relax}%
\providecommand \@@startlink[1]{}%
\providecommand \@@endlink[0]{}%
\providecommand \url  [0]{\begingroup\@sanitize@url \@url }%
\providecommand \@url [1]{\endgroup\@href {#1}{\urlprefix }}%
\providecommand \urlprefix  [0]{URL }%
\providecommand \Eprint [0]{\href }%
\providecommand \doibase [0]{http://dx.doi.org/}%
\providecommand \selectlanguage [0]{\@gobble}%
\providecommand \bibinfo  [0]{\@secondoftwo}%
\providecommand \bibfield  [0]{\@secondoftwo}%
\providecommand \translation [1]{[#1]}%
\providecommand \BibitemOpen [0]{}%
\providecommand \bibitemStop [0]{}%
\providecommand \bibitemNoStop [0]{.\EOS\space}%
\providecommand \EOS [0]{\spacefactor3000\relax}%
\providecommand \BibitemShut  [1]{\csname bibitem#1\endcsname}%
\let\auto@bib@innerbib\@empty
\bibitem [{\citenamefont {Grad}(1967)}]{grad1967toroidal}%
  \BibitemOpen
  \bibfield  {author} {\bibinfo {author} {\bibfnamefont {H.}~\bibnamefont
  {Grad}},\ }\href@noop {} {\bibfield  {journal} {\bibinfo  {journal} {The
  Physics of Fluids}\ }\textbf {\bibinfo {volume} {10}},\ \bibinfo {pages}
  {137} (\bibinfo {year} {1967})}\BibitemShut {NoStop}%
\bibitem [{\citenamefont {Newcomb}(1959)}]{newcomb1959magnetic}%
  \BibitemOpen
  \bibfield  {author} {\bibinfo {author} {\bibfnamefont {W.~A.}\ \bibnamefont
  {Newcomb}},\ }\href@noop {} {\bibfield  {journal} {\bibinfo  {journal} {The
  Physics of Fluids}\ }\textbf {\bibinfo {volume} {2}},\ \bibinfo {pages} {362}
  (\bibinfo {year} {1959})}\BibitemShut {NoStop}%
\bibitem [{\citenamefont {Hudson}\ and\ \citenamefont
  {Kraus}(2017)}]{hudsonKrauss20173D_cont_B}%
  \BibitemOpen
  \bibfield  {author} {\bibinfo {author} {\bibfnamefont {S.}~\bibnamefont
  {Hudson}}\ and\ \bibinfo {author} {\bibfnamefont {B.}~\bibnamefont {Kraus}},\
  }\href@noop {} {\bibfield  {journal} {\bibinfo  {journal} {Journal of Plasma
  Physics}\ }\textbf {\bibinfo {volume} {83}} (\bibinfo {year}
  {2017})}\BibitemShut {NoStop}%
\bibitem [{\citenamefont {Firpo}\ and\ \citenamefont
  {Constantinescu}(2011)}]{firpo2011study}%
  \BibitemOpen
  \bibfield  {author} {\bibinfo {author} {\bibfnamefont {M.-C.}\ \bibnamefont
  {Firpo}}\ and\ \bibinfo {author} {\bibfnamefont {D.}~\bibnamefont
  {Constantinescu}},\ }\href@noop {} {\bibfield  {journal} {\bibinfo  {journal}
  {Physics of Plasmas}\ }\textbf {\bibinfo {volume} {18}},\ \bibinfo {pages}
  {032506} (\bibinfo {year} {2011})}\BibitemShut {NoStop}%
\bibitem [{\citenamefont {Faber}\ \emph {et~al.}(2018)\citenamefont {Faber},
  \citenamefont {Pueschel}, \citenamefont {Terry}, \citenamefont {Hegna},\ and\
  \citenamefont {Roman}}]{faber2018stellarator}%
  \BibitemOpen
  \bibfield  {author} {\bibinfo {author} {\bibfnamefont {B.}~\bibnamefont
  {Faber}}, \bibinfo {author} {\bibfnamefont {M.}~\bibnamefont {Pueschel}},
  \bibinfo {author} {\bibfnamefont {P.}~\bibnamefont {Terry}}, \bibinfo
  {author} {\bibfnamefont {C.}~\bibnamefont {Hegna}}, \ and\ \bibinfo {author}
  {\bibfnamefont {J.}~\bibnamefont {Roman}},\ }\href@noop {} {\bibfield
  {journal} {\bibinfo  {journal} {Journal of Plasma Physics}\ }\textbf
  {\bibinfo {volume} {84}} (\bibinfo {year} {2018})}\BibitemShut {NoStop}%
\bibitem [{\citenamefont {Hirsch}\ \emph {et~al.}(2008)\citenamefont {Hirsch},
  \citenamefont {Baldzuhn}, \citenamefont {Beidler}, \citenamefont {Brakel},
  \citenamefont {Burhenn}, \citenamefont {Dinklage}, \citenamefont {Ehmler},
  \citenamefont {Endler}, \citenamefont {Erckmann}, \citenamefont {Feng},
  \citenamefont {Geiger}, \citenamefont {Giannone}, \citenamefont {Grieger},
  \citenamefont {Grigull}, \citenamefont {Hartfu{\ss}}, \citenamefont
  {Hartmann}, \citenamefont {Jaenicke}, \citenamefont {K\"onig}, \citenamefont
  {Laqua}, \citenamefont {Maa{\ss}berg}, \citenamefont {McCormick},
  \citenamefont {Sardei}, \citenamefont {Speth}, \citenamefont {Stroth},
  \citenamefont {Wagner}, \citenamefont {Weller}, \citenamefont {Werner},
  \citenamefont {Wobig}, \citenamefont {Zoletnik},\ and\ \citenamefont {the
  {W7-AS Team}}}]{hirsch2008majorW7AS}%
  \BibitemOpen
  \bibfield  {author} {\bibinfo {author} {\bibfnamefont {M.}~\bibnamefont
  {Hirsch}}, \bibinfo {author} {\bibfnamefont {J.}~\bibnamefont {Baldzuhn}},
  \bibinfo {author} {\bibfnamefont {C.}~\bibnamefont {Beidler}}, \bibinfo
  {author} {\bibfnamefont {R.}~\bibnamefont {Brakel}}, \bibinfo {author}
  {\bibfnamefont {R.}~\bibnamefont {Burhenn}}, \bibinfo {author} {\bibfnamefont
  {A.}~\bibnamefont {Dinklage}}, \bibinfo {author} {\bibfnamefont
  {H.}~\bibnamefont {Ehmler}}, \bibinfo {author} {\bibfnamefont
  {M.}~\bibnamefont {Endler}}, \bibinfo {author} {\bibfnamefont
  {V.}~\bibnamefont {Erckmann}}, \bibinfo {author} {\bibfnamefont
  {Y.}~\bibnamefont {Feng}}, \bibinfo {author} {\bibfnamefont {J.}~\bibnamefont
  {Geiger}}, \bibinfo {author} {\bibfnamefont {L.}~\bibnamefont {Giannone}},
  \bibinfo {author} {\bibfnamefont {G.}~\bibnamefont {Grieger}}, \bibinfo
  {author} {\bibfnamefont {P.}~\bibnamefont {Grigull}}, \bibinfo {author}
  {\bibfnamefont {H.-J.}\ \bibnamefont {Hartfu{\ss}}}, \bibinfo {author}
  {\bibfnamefont {D.}~\bibnamefont {Hartmann}}, \bibinfo {author}
  {\bibfnamefont {R.}~\bibnamefont {Jaenicke}}, \bibinfo {author}
  {\bibfnamefont {R.}~\bibnamefont {K\"onig}}, \bibinfo {author} {\bibfnamefont
  {H.~P.}\ \bibnamefont {Laqua}}, \bibinfo {author} {\bibfnamefont
  {H.}~\bibnamefont {Maa{\ss}berg}}, \bibinfo {author} {\bibfnamefont
  {K.}~\bibnamefont {McCormick}}, \bibinfo {author} {\bibfnamefont
  {F.}~\bibnamefont {Sardei}}, \bibinfo {author} {\bibfnamefont
  {E.}~\bibnamefont {Speth}}, \bibinfo {author} {\bibfnamefont
  {U.}~\bibnamefont {Stroth}}, \bibinfo {author} {\bibfnamefont
  {F.}~\bibnamefont {Wagner}}, \bibinfo {author} {\bibfnamefont
  {A.}~\bibnamefont {Weller}}, \bibinfo {author} {\bibfnamefont
  {A.}~\bibnamefont {Werner}}, \bibinfo {author} {\bibfnamefont
  {H.}~\bibnamefont {Wobig}}, \bibinfo {author} {\bibfnamefont
  {S.}~\bibnamefont {Zoletnik}}, \ and\ \bibinfo {author} {\bibnamefont {the
  {W7-AS Team}}},\ }\href@noop {} {\bibfield  {journal} {\bibinfo  {journal}
  {Plasma Physics and Controlled Fusion}\ }\textbf {\bibinfo {volume} {50}}
  (\bibinfo {year} {2008})}\BibitemShut {NoStop}%
\bibitem [{\citenamefont {Brakel}\ and\ \citenamefont {the
  {W7-AS}~Team}(2002)}]{brakel2002energytransp_rational_iota_W7AS}%
  \BibitemOpen
  \bibfield  {author} {\bibinfo {author} {\bibfnamefont {R.}~\bibnamefont
  {Brakel}}\ and\ \bibinfo {author} {\bibnamefont {the {W7-AS}~Team}},\
  }\href@noop {} {\bibfield  {journal} {\bibinfo  {journal} {Nuclear fusion}\
  }\textbf {\bibinfo {volume} {42}},\ \bibinfo {pages} {903} (\bibinfo {year}
  {2002})}\BibitemShut {NoStop}%
\bibitem [{\citenamefont {Brakel}\ \emph {et~al.}(1997)\citenamefont {Brakel},
  \citenamefont {Anton}, \citenamefont {Baldzuhn}, \citenamefont {Burhenn},
  \citenamefont {Erckmann}, \citenamefont {Fiedler}, \citenamefont {Geiger},
  \citenamefont {Hartfuss}, \citenamefont {Heinrich}, \citenamefont {Hirsch},
  \citenamefont {Jaenicke}, \citenamefont {Kick}, \citenamefont {K\"uhner},
  \citenamefont {Maa{\ss}berg}, \citenamefont {Stroth}, \citenamefont {Wagner},
  \citenamefont {Weller}, \citenamefont {{W7-AS Team}}, \citenamefont {{ECRH
  Group}},\ and\ \citenamefont {{NBI-Group}}}]{brakel1997W7AS_EB_shear}%
  \BibitemOpen
  \bibfield  {author} {\bibinfo {author} {\bibfnamefont {R.}~\bibnamefont
  {Brakel}}, \bibinfo {author} {\bibfnamefont {M.}~\bibnamefont {Anton}},
  \bibinfo {author} {\bibfnamefont {J.}~\bibnamefont {Baldzuhn}}, \bibinfo
  {author} {\bibfnamefont {R.}~\bibnamefont {Burhenn}}, \bibinfo {author}
  {\bibfnamefont {V.}~\bibnamefont {Erckmann}}, \bibinfo {author}
  {\bibfnamefont {S.}~\bibnamefont {Fiedler}}, \bibinfo {author} {\bibfnamefont
  {J.}~\bibnamefont {Geiger}}, \bibinfo {author} {\bibfnamefont
  {H.}~\bibnamefont {Hartfuss}}, \bibinfo {author} {\bibfnamefont
  {O.}~\bibnamefont {Heinrich}}, \bibinfo {author} {\bibfnamefont
  {M.}~\bibnamefont {Hirsch}}, \bibinfo {author} {\bibfnamefont
  {R.}~\bibnamefont {Jaenicke}}, \bibinfo {author} {\bibfnamefont
  {M.}~\bibnamefont {Kick}}, \bibinfo {author} {\bibfnamefont {G.}~\bibnamefont
  {K\"uhner}}, \bibinfo {author} {\bibfnamefont {H.}~\bibnamefont
  {Maa{\ss}berg}}, \bibinfo {author} {\bibfnamefont {U.}~\bibnamefont
  {Stroth}}, \bibinfo {author} {\bibfnamefont {F.}~\bibnamefont {Wagner}},
  \bibinfo {author} {\bibfnamefont {A.}~\bibnamefont {Weller}}, \bibinfo
  {author} {\bibnamefont {{W7-AS Team}}}, \bibinfo {author} {\bibnamefont
  {{ECRH Group}}}, \ and\ \bibinfo {author} {\bibnamefont {{NBI-Group}}},\
  }\href@noop {} {\bibfield  {journal} {\bibinfo  {journal} {Plasma Physics and
  Controlled Fusion}\ }\textbf {\bibinfo {volume} {39}},\ \bibinfo {pages}
  {B273} (\bibinfo {year} {1997})}\BibitemShut {NoStop}%
\bibitem [{\citenamefont {Wobig}(1987)}]{wobig1987localized_pert_w7A}%
  \BibitemOpen
  \bibfield  {author} {\bibinfo {author} {\bibfnamefont {H.}~\bibnamefont
  {Wobig}},\ }\href@noop {} {\bibfield  {journal} {\bibinfo  {journal}
  {Zeitschrift f{\"u}r Naturforschung A}\ }\textbf {\bibinfo {volume} {42}},\
  \bibinfo {pages} {1054} (\bibinfo {year} {1987})}\BibitemShut {NoStop}%
\bibitem [{\citenamefont {Andreeva}(2002)}]{andreevaW7Xvacuum}%
  \BibitemOpen
  \bibfield  {author} {\bibinfo {author} {\bibfnamefont {T.}~\bibnamefont
  {Andreeva}},\ }\href@noop {} {\enquote {\bibinfo {title} {Vacuum magnetic
  configurations of wendelstein 7-x},}\ }\bibinfo {type} {Tech. Rep.}\
  (\bibinfo  {institution} {Max-Planck-Institut fuer Plasmaphysik},\ \bibinfo
  {year} {2002})\BibitemShut {NoStop}%
\bibitem [{\citenamefont {Grad}(1973)}]{grad1973magnetofluid}%
  \BibitemOpen
  \bibfield  {author} {\bibinfo {author} {\bibfnamefont {H.}~\bibnamefont
  {Grad}},\ }\href@noop {} {\bibfield  {journal} {\bibinfo  {journal}
  {Proceedings of the National Academy of Sciences}\ }\textbf {\bibinfo
  {volume} {70}},\ \bibinfo {pages} {3277} (\bibinfo {year}
  {1973})}\BibitemShut {NoStop}%
\bibitem [{\citenamefont {Strauss}\ and\ \citenamefont
  {Monticello}(1981)}]{strauss1981limiting_beta_vacuum}%
  \BibitemOpen
  \bibfield  {author} {\bibinfo {author} {\bibfnamefont {H.}~\bibnamefont
  {Strauss}}\ and\ \bibinfo {author} {\bibfnamefont {D.}~\bibnamefont
  {Monticello}},\ }\href@noop {} {\bibfield  {journal} {\bibinfo  {journal}
  {The Physics of Fluids}\ }\textbf {\bibinfo {volume} {24}},\ \bibinfo {pages}
  {1148} (\bibinfo {year} {1981})}\BibitemShut {NoStop}%
\bibitem [{\citenamefont {Klinger}\ \emph {et~al.}(2019)\citenamefont
  {Klinger}, \citenamefont {Andreeva}, \citenamefont {Bozhenkov}, \citenamefont
  {Brandt}, \citenamefont {Burhenn}, \citenamefont {Buttensch{\"o}n},
  \citenamefont {Fuchert}, \citenamefont {Geiger}, \citenamefont {Grulke},
  \citenamefont {Laqua} \emph {et~al.}}]{klinger2019overview}%
  \BibitemOpen
  \bibfield  {author} {\bibinfo {author} {\bibfnamefont {T.}~\bibnamefont
  {Klinger}}, \bibinfo {author} {\bibfnamefont {T.}~\bibnamefont {Andreeva}},
  \bibinfo {author} {\bibfnamefont {S.}~\bibnamefont {Bozhenkov}}, \bibinfo
  {author} {\bibfnamefont {C.}~\bibnamefont {Brandt}}, \bibinfo {author}
  {\bibfnamefont {R.}~\bibnamefont {Burhenn}}, \bibinfo {author} {\bibfnamefont
  {B.}~\bibnamefont {Buttensch{\"o}n}}, \bibinfo {author} {\bibfnamefont
  {G.}~\bibnamefont {Fuchert}}, \bibinfo {author} {\bibfnamefont
  {B.}~\bibnamefont {Geiger}}, \bibinfo {author} {\bibfnamefont
  {O.}~\bibnamefont {Grulke}}, \bibinfo {author} {\bibfnamefont
  {H.}~\bibnamefont {Laqua}},  \emph {et~al.},\ }\href@noop {} {\bibfield
  {journal} {\bibinfo  {journal} {Nuclear Fusion}\ }\textbf {\bibinfo {volume}
  {59}},\ \bibinfo {pages} {112004} (\bibinfo {year} {2019})}\BibitemShut
  {NoStop}%
\bibitem [{\citenamefont {Lazerson}\ \emph {et~al.}(2019)\citenamefont
  {Lazerson}, \citenamefont {Gao}, \citenamefont {Hammond}, \citenamefont
  {Killer}, \citenamefont {Schlisio}, \citenamefont {Otte}, \citenamefont
  {Biedermann}, \citenamefont {Spolaore}, \citenamefont {Bozhenkov},
  \citenamefont {Geiger} \emph {et~al.}}]{lazerson2019tuning}%
  \BibitemOpen
  \bibfield  {author} {\bibinfo {author} {\bibfnamefont {S.~A.}\ \bibnamefont
  {Lazerson}}, \bibinfo {author} {\bibfnamefont {Y.}~\bibnamefont {Gao}},
  \bibinfo {author} {\bibfnamefont {K.}~\bibnamefont {Hammond}}, \bibinfo
  {author} {\bibfnamefont {C.}~\bibnamefont {Killer}}, \bibinfo {author}
  {\bibfnamefont {G.}~\bibnamefont {Schlisio}}, \bibinfo {author}
  {\bibfnamefont {M.}~\bibnamefont {Otte}}, \bibinfo {author} {\bibfnamefont
  {C.}~\bibnamefont {Biedermann}}, \bibinfo {author} {\bibfnamefont
  {M.}~\bibnamefont {Spolaore}}, \bibinfo {author} {\bibfnamefont
  {S.}~\bibnamefont {Bozhenkov}}, \bibinfo {author} {\bibfnamefont
  {J.}~\bibnamefont {Geiger}},  \emph {et~al.},\ }\href@noop {} {\bibfield
  {journal} {\bibinfo  {journal} {Nuclear Fusion}\ }\textbf {\bibinfo {volume}
  {59}},\ \bibinfo {pages} {126004} (\bibinfo {year} {2019})}\BibitemShut
  {NoStop}%
\bibitem [{\citenamefont {Lortz}(1970)}]{lortz1970existenz}%
  \BibitemOpen
  \bibfield  {author} {\bibinfo {author} {\bibfnamefont {D.}~\bibnamefont
  {Lortz}},\ }\href@noop {} {\bibfield  {journal} {\bibinfo  {journal}
  {Zeitschrift f{\"u}r angewandte Mathematik und Physik ZAMP}\ }\textbf
  {\bibinfo {volume} {21}},\ \bibinfo {pages} {196} (\bibinfo {year}
  {1970})}\BibitemShut {NoStop}%
\bibitem [{\citenamefont {Cary}(1982)}]{cary1982vacuum}%
  \BibitemOpen
  \bibfield  {author} {\bibinfo {author} {\bibfnamefont {J.~R.}\ \bibnamefont
  {Cary}},\ }\href@noop {} {\bibfield  {journal} {\bibinfo  {journal} {Physical
  Review Letters}\ }\textbf {\bibinfo {volume} {49}},\ \bibinfo {pages} {276}
  (\bibinfo {year} {1982})}\BibitemShut {NoStop}%
\bibitem [{\citenamefont {Freidberg}(1982)}]{freidberg_idealMHD}%
  \BibitemOpen
  \bibfield  {author} {\bibinfo {author} {\bibfnamefont {J.~P.}\ \bibnamefont
  {Freidberg}},\ }\href@noop {} {\bibfield  {journal} {\bibinfo  {journal}
  {Reviews of Modern Physics}\ }\textbf {\bibinfo {volume} {54}},\ \bibinfo
  {pages} {801} (\bibinfo {year} {1982})}\BibitemShut {NoStop}%
\bibitem [{\citenamefont {Sengupta}\ and\ \citenamefont
  {Weitzner}(2018)}]{sengupta_weitzner2018}%
  \BibitemOpen
  \bibfield  {author} {\bibinfo {author} {\bibfnamefont {W.}~\bibnamefont
  {Sengupta}}\ and\ \bibinfo {author} {\bibfnamefont {H.}~\bibnamefont
  {Weitzner}},\ }\href@noop {} {\bibfield  {journal} {\bibinfo  {journal}
  {Physics of Plasmas}\ }\textbf {\bibinfo {volume} {25}},\ \bibinfo {pages}
  {022506} (\bibinfo {year} {2018})}\BibitemShut {NoStop}%
\bibitem [{\citenamefont {Weitzner}(2014)}]{weitzner2014ideal}%
  \BibitemOpen
  \bibfield  {author} {\bibinfo {author} {\bibfnamefont {H.}~\bibnamefont
  {Weitzner}},\ }\href@noop {} {\bibfield  {journal} {\bibinfo  {journal}
  {Physics of Plasmas}\ }\textbf {\bibinfo {volume} {21}},\ \bibinfo {pages}
  {022515} (\bibinfo {year} {2014})}\BibitemShut {NoStop}%
\bibitem [{\citenamefont {Cary}(1984)}]{cary1984construction}%
  \BibitemOpen
  \bibfield  {author} {\bibinfo {author} {\bibfnamefont {J.~R.}\ \bibnamefont
  {Cary}},\ }\href@noop {} {\bibfield  {journal} {\bibinfo  {journal} {The
  Physics of fluids}\ }\textbf {\bibinfo {volume} {27}},\ \bibinfo {pages}
  {119} (\bibinfo {year} {1984})}\BibitemShut {NoStop}%
\bibitem [{\citenamefont {Mercier}(1964)}]{Mercier1964}%
  \BibitemOpen
  \bibfield  {author} {\bibinfo {author} {\bibfnamefont {C.}~\bibnamefont
  {Mercier}},\ }\href {\doibase 10.1088/0029-5515/4/3/008} {\bibfield
  {journal} {\bibinfo  {journal} {Nuclear Fusion}\ }\textbf {\bibinfo {volume}
  {4}},\ \bibinfo {pages} {213} (\bibinfo {year} {1964})}\BibitemShut {NoStop}%
\bibitem [{\citenamefont {Solov'ev}\ and\ \citenamefont
  {Shafranov}(1970)}]{Solovev1970}%
  \BibitemOpen
  \bibfield  {author} {\bibinfo {author} {\bibfnamefont {L.~S.}\ \bibnamefont
  {Solov'ev}}\ and\ \bibinfo {author} {\bibfnamefont {V.~D.}\ \bibnamefont
  {Shafranov}},\ }\href@noop {} {\emph {\bibinfo {title} {{Reviews of Plasma
  Physics 5}}}}\ (\bibinfo  {publisher} {Consultants Bureau},\ \bibinfo
  {address} {New York - London},\ \bibinfo {year} {1970})\BibitemShut {NoStop}%
\bibitem [{\citenamefont {Weitzner}(2016)}]{weitzner2016expansions}%
  \BibitemOpen
  \bibfield  {author} {\bibinfo {author} {\bibfnamefont {H.}~\bibnamefont
  {Weitzner}},\ }\href@noop {} {\bibfield  {journal} {\bibinfo  {journal}
  {Physics of Plasmas}\ }\textbf {\bibinfo {volume} {23}},\ \bibinfo {pages}
  {062512} (\bibinfo {year} {2016})}\BibitemShut {NoStop}%
\bibitem [{\citenamefont {Sengupta}\ and\ \citenamefont
  {Weitzner}(2019)}]{sengupta_weitzner_2019}%
  \BibitemOpen
  \bibfield  {author} {\bibinfo {author} {\bibfnamefont {W.}~\bibnamefont
  {Sengupta}}\ and\ \bibinfo {author} {\bibfnamefont {H.}~\bibnamefont
  {Weitzner}},\ }\href {\doibase 10.1017/S0022377819000230} {\bibfield
  {journal} {\bibinfo  {journal} {Journal of Plasma Physics}\ }\textbf
  {\bibinfo {volume} {85}},\ \bibinfo {pages} {905850209} (\bibinfo {year}
  {2019})}\BibitemShut {NoStop}%
\bibitem [{\citenamefont {Landreman}\ and\ \citenamefont
  {Sengupta}(2018)}]{landreman_Sengupta2018direct}%
  \BibitemOpen
  \bibfield  {author} {\bibinfo {author} {\bibfnamefont {M.}~\bibnamefont
  {Landreman}}\ and\ \bibinfo {author} {\bibfnamefont {W.}~\bibnamefont
  {Sengupta}},\ }\href@noop {} {\bibfield  {journal} {\bibinfo  {journal}
  {Journal of Plasma Physics}\ }\textbf {\bibinfo {volume} {84}} (\bibinfo
  {year} {2018})}\BibitemShut {NoStop}%
\bibitem [{\citenamefont {Landreman}, \citenamefont {Sengupta},\ and\
  \citenamefont {Plunk}(2019)}]{landreman_Sengupta_Plunck2019direct}%
  \BibitemOpen
  \bibfield  {author} {\bibinfo {author} {\bibfnamefont {M.}~\bibnamefont
  {Landreman}}, \bibinfo {author} {\bibfnamefont {W.}~\bibnamefont {Sengupta}},
  \ and\ \bibinfo {author} {\bibfnamefont {G.~G.}\ \bibnamefont {Plunk}},\
  }\href@noop {} {\bibfield  {journal} {\bibinfo  {journal} {Journal of Plasma
  Physics}\ }\textbf {\bibinfo {volume} {85}} (\bibinfo {year}
  {2019})}\BibitemShut {NoStop}%
\bibitem [{\citenamefont {Grad}(1971)}]{grad1971plasma}%
  \BibitemOpen
  \bibfield  {author} {\bibinfo {author} {\bibfnamefont {H.}~\bibnamefont
  {Grad}},\ }in\ \href@noop {} {\emph {\bibinfo {booktitle} {Plasma Physics and
  Controlled Nuclear Fusion Research 1971. Vol. III. Proceedings of the Fourth
  International Conference on Plasma Physics and Controlled Nuclear Fusion
  Research}}}\ (\bibinfo {year} {1971})\BibitemShut {NoStop}%
\end{thebibliography}%

\end{document}